\documentclass[review]{elsarticle}

\usepackage{lineno,hyperref}
\usepackage{threeparttable}
\usepackage{pdflscape}
\usepackage{float}
\usepackage{amssymb}
\usepackage{amsmath}

\usepackage{verbatim}
\usepackage{comment}
\newcommand{\comm}[1]{}
\usepackage{color}
\usepackage{cancel}
\usepackage{amstext}

\usepackage{bm}
\usepackage{enumerate}
\usepackage{xr-hyper}
\usepackage{cleveref}
\journal{ }

\begin{document}

\begin{frontmatter}

\title{Mechanical stability conditions for 3D and 2D crystals under arbitrary load}

\author[]{Marcin Ma\'zdziarz\corref{mycorrespondingauthor}}
\ead{mmazdz@ippt.pan.pl}

\address{Institute of Fundamental Technological Research Polish Academy of Sciences,
Pawi\'nskiego 5B, 02-106 Warsaw, Poland}
\cortext[mycorrespondingauthor]{Corresponding author}

\begin{abstract}
{The paper gathers and unifies mechanical stability conditions for all symmetry classes of 3D and 2D materials under arbitrary load. The methodology is based on the spectral decomposition of the {fourth-order} stiffness tensors mapped to {second-order} tensors using \emph{orthonormal} (Mandel) notation, and the verification of the positivity of the so-called Kelvin moduli. An explicit set of stability conditions for 3D and 2D crystals of higher symmetry is also included, as well as a Mathematica notebook that allows mechanical stability analysis for crystals, stress-free and stressed, of arbitrary symmetry under arbitrary loads.}
\end{abstract}

\begin{keyword}
	Mechanical stability\sep
	Born's stability\sep
	2D materials\sep
	Kelvin moduli\sep
	orthonormal notation
\end{keyword}

\end{frontmatter}



\section{Introduction}
\label{sec:Int}

The examination of the mechanical stability of crystals is inextricably linked to the seminal paper of Max Born from 1940, in which he analyzed the stability of unstressed crystals with cubic symmetry \cite{Born1940}. This stability is often referred to as Born's stability. Stability conditions for unstressed 3D crystals of arbitrary symmetry can be found, for example, in \cite{Mouhat2014}. Unfortunately, it uses non-tensorial Voigt notation and the principal minors of the stiffness tensor, making these conditions non-objective and thus dependent on the orientation of the crystal.

In general, checking the mechanical stability boils down to checking the positive definiteness of a certain {fourth-order} stiffness tensor, i.e., the non-negativity of a certain quadratic form for arbitrary values of incremental strains \cite{Morris2000}. Historically, and especially in the field of continuum mechanics, there are at least four stability criteria for the loaded material, which differ in the choice of appropriate stiffness tensors \cite{Clayton2012, Jog2013}. We denote these criteria by:\comm{\textbf{C}} $\mathbb{C}$, where the stiffness tensor is based on a symmetric Green strain; \comm{\textbf{A}} $\mathbb{A}$, where the stiffness tensor is based on an asymmetric deformation gradient; \comm{\textbf{Z}} $\mathbb{Z}$, like \comm{\textbf{A}} $\mathbb{A}$ except that in the absence of rotation and the incremental stiffness tensor \comm{\textbf{B}} $\mathbb{L}$ giving change in Cauchy stress with respect to strain. In the  \emph{ab\,initio} and molecular calculations, only the $\mathbb{L}$ and $\mathbb{C}$ criteria are relevant. Due to the polymorphism of the crystals, the stress-free configuration is not uniquely determined and only the current configuration is well defined. These criteria are also the most and least stringent \cite{Clayton2014}. 
\comm{Of practical importance, especially in \emph{ab\,initio} and molecular calculations, criteria \comm{\textbf{B}} $\mathbb{L}$ and \comm{\textbf{C}} $\mathbb{C}$ are the most and least stringent \cite{Clayton2014}, respectively.} If the crystal is in any stress-free configuration, then all of these criteria are equivalent.  

The phonon (dynamical)- and mechanical (Born)-stability criteria are, in a sense, complementary and must be satisfied \cite{ELLIOTT2006161}. There are 3 acoustic branches of the phonon dispersion relation that start linearly at $\Gamma$ point. The slope of the dispersion curve near $\Gamma$ is the speed of sound. Knowing these slopes, the point symmetry of the crystal and its density, we can deduce the full stiffness tensor. The dynamical stability condition requires that all phonon modes have positive frequencies. Positive slopes of acoustic branches at $\Gamma$ point do not imply mechanical stability \cite{Cowley1976}, these conditions are more complex.

When checking mechanical stability, the problem is not that its conditions are not known, but that they are defined by {fourth-order} tensors and cannot be applied directly, they must be transformed into a useful matrix or lower-order tensor form. Even in this transformed form, for crystals of lower symmetry, this verification can be very cumbersome \cite{Mouhat2014}. For the \comm{\textbf{B}} $\mathbb{L}$ criterion, only special, simplest cases are discussed in the literature, i.e. cubic crystals under hydrostatic pressure loading \cite{Sinko2002}, uniaxial tensile deformation \cite{Wen2017} and simple shear \cite{Morris2000}, respectively.
The methodology proposed here does not require the checking of a list of conditions. Using the spectral decomposition of the fourth-order stiffness tensors mapped to a second-order tensors using \emph{orthonormal} (Mandel) notation, for any material symmetry and any loading, amounts to computing Kelvin moduli, i.e., appropriate eigenvalues. The procedure requires only writing the calculated, either \textit{ab\,initio} or atomistic, elastic constants and components of the stress tensor into the proper form of second-order tensors using \emph{orthonormal} notation, and then solving the eigenproblem numerically. 

The remainder of this paper is structured as follows: in Sec.\ref{sec:srtn} the differences between Voigt and \emph{orthonormal} notations are discussed in detail,
in Sec.\ref{sec:Isc} general mechanical stability conditions for stress-free systems of arbitrary symmetry written in \emph{orthonormal} notation are discussed and 
in Sec.\ref{sec:Iscds} the above stability conditions for stress-free systems were generalized to deformed and stressed systems.  
Additionally in \comm{Appendix} \ref{sec:AppA} representations of the stiffness tensor for stress-free B2 NiAl written in \emph{orthonormal} notation for three different orientations are shown, 
in \comm{Appendix} \ref{sec:AppB} the application of general stability conditions to the cubic B2 NiAl crystal subjected to biaxial deformation, {tension and compression,} is presented. 
To complete the paper, an explicit set of stability conditions using both leading principal minors and Kelvin moduli is given in \comm{Appendix} \ref{sec:App3D} for stress-free 3D systems with standard lattice vectors, and in \comm{Appendix} \ref{sec:App2D} for 2D systems. Explicit formulas for homogenized isotropic bulk and shear modulus are given in \comm{Appendix} \ref{sec:AppE}.

\comm
{
mozna tez  rozwazyc PHYSICAL REVIEW LETTERS 200p "born stability" 14 wystapien

:( Materials Science and Engineering B https://www.sciencedirect.com/journal/materials-science-and-engineering-b 100p, IF=3.9 born stability 93 pozycje
:( Journal of Computational Physics https://www.sciencedirect.com/journal/journal-of-computational-physics 140p, IF=3.8 born stability conditions 124 pozycje
? SciPost Physics https://scipost.org/SciPostPhys 100p, IF=4.6 OA free
PROCEEDINGS OF THE ROYAL SOCIETY A-MATHEMATICAL PHYSICAL AND ENGINEERING SCIENCES 100p, IF=2.9, 47 results for""mechanical stability""
Philosophical Transactions of the Royal Society A: Mathematical, Physical and Engineering Sciences 100p, IF=4.3  67 results for""mechanical stability""
:( Journal of Materials Research and Technology https://www.sciencedirect.com/journal/journal-of-materials-research-and-technology 100p, IF=6.2 OA 
:( Computational Materials Science, "born stability" - 109 results, "mechanical stability"-707 results
IEEE Transactions on Nanotechnology https://ieeexplore.ieee.org/xpl/RecentIssue.jsp?punumber=7729 100p, IF=2.1

Moze i to zacytowac? \cite{Jog2013} "{Conditions for the onset of elastic and material instabilities in hyperelastic materials}" For an isotropic material at F = I , the necessary and sufficient conditions for strong ellipticity are $\mu$ > 0 and $\lambda$ + 2$\mu$ > 0. Thus, the bulk modulus $\kappa$ := $\lambda$ + 2$\mu$/3 need not be positive, which is an important physical
requirement that ensures that a sphere ‘compresses’ when subjected to pressure loading on its surface.

\cite{Bukharaev_2018} "{Straintronics: a new trend in micro- and nanoelectronics and materials science}", The term 'straintronics' refers to a new research area in condensed matter physics, in which strain engineering
methods and strain-induced physical effects in solids are used to develop next-generation devices for information, sensor, and energy-saving technologies.
	
Problemy z warunkami z \cite{Mouhat2014, ting1992anisotropic, Gao2023} : Voigt notation, bazuja na the leading principal minors of C complexity is $O$(n$^3$) , warunki sa wazne tylko dla jednej bazy wektorow sieci nie sa ogolne , dotyczy to tez warunkow dla symetrii kubicznej z \cite{Born1940}

{ \cite{ELLIOTT2006161} "{Stability of crystalline solids—I: Continuum and atomic lattice considerations}" There is overlap between the phonon-stability and CB-stability criteria. If one
considers perturbations with no uniform deformation component, dU  0, then the components q2~E=qs2 of the CB-stability matrix, Eq. (2.64) (in the limit as n!1, for the infinite crystal), are positive definite only if the optic phonon equations (2.53) have positive eigenvalues, . 
The phonon (dynamical)- and mechanical (Born)-stability criteria are, in a sense, complementary. There are 3 acoustic branches of the phonon dispersion relation that start linearly at $\Gamma$ point. The slope of the dispersion curve near $\Gamma$ is the speed of sound. Knowing these slopes, the point symmetry of the crystal and its density, we can deduce the full stiffness tensor. The dynamical stability condition requires that all phonon modes have positive frequencies. Positive slopes of acoustic branches at $\Gamma$ point do not imply mechanical stability \cite{Cowley1976}, these conditions are more complicated..... Part II "the
CB-stability criterion and the phonon-stability criterion must be satisfied. This ensures
stability with respect to all quasi-uniform perturbations (CB) (3.7) and bounded displacement
perturbations of all wavelengths (phonon)"

$https://www.ctcms.nist.gov/potentials/atomman/tutorial/3.1._ElasticConstants_class.html$

Uzywaja powszechnie {Voigt} notation \cite{Sinko2002}-(stabilnosc tylko pod cisnieniem) 

\cite{Mouhat2014} \cite{ting1992anisotropic}-the condition for positivity of all the principal minors


\cite{Gao2023} "Elastic stability criteria of seven crystal systems and their application under pressure: Taking carbon as an example", Void notation, for low-symmetry systems warunki sa cumbersome i wyprowadzenie jest prawdziwe tylko okreslonej reprezentacji Cij czyli only for a specific orientation of the crystallographic axes ... "all the leading principal minors of C (determinants of its upper-left k by k sub-matrix, 1 $\leq$ k $\leq$ 6) are positive." , complexity is $O$(n$^3$)

\cite{SINGH2021108068} "MechElastic: A Python Library for Analysis of Mechanical and Elastic Properties of Bulk and 2D Materials" - cytuje mnie, Voigt notation
\cite{LIU2022108180} "ElasTool: An automated toolkit for elastic constants calculation" -> Voigt notation
}
and similarly under hydrostatic pressure in \cite{Grimvall2012} "Lattice instabilities in metallic elements" Opis wykorzystac: Here Cijkl are elements in the elastic constant tensor C evaluated at the current stressed state (which may not have cubic symmetry), $\sigma_{ij}$ specify the external stresses

Fajna praca \cite{Chadwick1971} "{On the definition of elastic moduli}"

Do zacytowania \cite{rychlewski2023ceiiinosssttuv} "{"Ceiiinosssttuv": Mathematical Structure of Elastic Bodies}"
Przydatne \cite{KO2009note}
tez przydatne \cite{Moakher2006} "{The Closest Elastic Tensor of Arbitrary Symmetry to an Elasticity Tensor of Lower Symmetry}"

Ciekawe \cite{Bigoni2000} "Bifurcation and Instability of Non-Associative Elastoplastic Solids" str 25 (93) pokazano ze positive definiteness of C (PD condition) implies strong ellipticity (SE)

\cite{CHADWICK20012471} "A new proof that the number of linear elastic symmetries is eight"

Prawie wszystko co mi potrzeba dla 3D \cite{Cowin1995} "Anisotropic symmetries of linear elasticity" wzor (15)

Moze przydatne \cite{Bona2007} "Coordinate-free Characterization of the Symmetry Classes of Elasticity Tensors": Fig. 1 Diagram representing relations among the symmetry classes

\comm{Warto moze douczyc sie Harmonic and Cartan decompositions \cite{Forte1996} "{Symmetry classes for elasticity tensors}" i \cite{Wen2024} "{An equivariant graph neural network for the elasticity tensors of all seven crystal systems}"
https://github.com/wengroup/matten?tab=readme-ov-file

Sprawdzic co naprawde jest liczone w ABINIT i porownac to z elasticity-oganov.pdf i  https://docs.abinit.org/tutorial/elastic/index.html 

Autorzy \cite{Nielsen1983} "{First-Principles Calculation of Stress}" licza virial stress, ktory odpowiada Cauchy stress Zdaja sobie sprawe z istnienia skonczonych deformacji, ale przeliczaja male na duze tylko dla jednowymiarowego przypadku

Fajne \cite{FEA2017} "Nonlinear finite elements$/$Rate form of hyperelastic laws" 
	
	W \cite{Nagel2016} "On advantages of the Kelvin mapping in finite element implementations of deformation processes"  ciekawa Table 1 Comparison of different tensor norms

	Dla 3D przepisac z \cite{BoehlkeBrueggemann2001} "Graphical Representation of the Generalized Hooke's Law", przepisac (8), (9)-(16)
	
	Przydatne \cite{KO2009note} i "Waves in Nonlinear Pre-Stressed Materials" rozdział I

pomysl jak u mnie tylko Voigt notation and uniaxial stress \cite{Wen2017} "{Lattice stability and the effect of Co and Re on the ideal strength of Ni: First-principles study of uniaxial tensile deformation}": To test the mechanical stability of a lattice during the tensile processes, we calculate the elastic constants Cij of the lattice at every tensile strain and examine the mechanical stability
conditions for the corresponding crystal according to Eqs. (9), (13), and (15). ... czy wzor (4) jak u mnie? 
}

Dla 3D "Table 3. Classification of crystal systems, point-group symbols and space-group number (S.G.N) are provided with the number of independent elastic constants for 3D materials. The material prototypes are shown in the last column." i 2D "Table 4. Classification of crystal systems, 2D space groups and the independent elastic constants are provided with the number of independent elastic constants for 2D materials. The material prototypes are shown in the last column." w AELAS \cite{ZHANG2017403} (Table 5 lattice vectors for 3D materials i Table 6 lattice vectors for 2D materials) Podzial na trig I i trig II, oraz tetra I i tetra II jest bledny \cite{Cowin1995}, wszystkie maja 6 stałych sprezystosci trzeba tylko dokonac wlasciwego podstawienia, tu tez blad \cite{Mouhat2014} It is also shown that the restrictions on the elastic coefficients appearing in Hooke's
law follow in a simple and straightforward fashion from orthogonal transformations based on a
small subset of the small catalogue of planes of mirror symmetry-Cowin

\begin{table}[H] 
	\begin{ruledtabular}
	\caption{The distinct symmetries of anisotropic elasticity and crystal systems for 3D materials.}
	\label{tab:3DS}
	\centering
	\tiny 
	\begin{tabular}{c c c c c c }
		Material  & Crystal  &  {Point } & Space  & No. of independent & No. of distinct  \\
		symmetry &  system &  group &  group & elastic constants & Kelvin moduli \\
		\hline Triclinic & Triclinic & C$_1$, C$_i$ & 1-2 & 21 & 6 \\
		Monoclinic & Monoclinic & C$_2$, C$_s$, C$_{2h}$ & 3–15 & 13 & 6 \\
		Orthotropic & Orthorhombic & D$_2$, C$_{2v}$, D$_{2h}$ & 16-74 & 9 & 6\\
		Tetragonal & Tetragonal & C$_4$, S$_4$, C$_{4h}$, D$_4$ & 75–142 & 6 & 5\\
		&  & C$_{4v}$, D$_{2d}$, D$_{4h}$ & &  \\
		Trigonal & Trigonal & C$_3$, S$_6$, D$_3$, C$_{3v}$, D$_{3d}$ & 143–167 & 6 & 4\\
		Transverse Isotropy & Hexagonal & C$_6$, C$_{3h}$, C$_{6h}$, D$_6$ & 168–194 & 5 & 4\\
		&  & C$_{6v}$, D$_{3h}$, D$_{6h}$ & &  \\
		Cubic & Cubic & T, T$_h$, O, T$_d$, O$_h$ & 195-230 & 3 & 3\\
		Isotropy & & &  & 2 & 2 
		\end{tabular}
\end{ruledtabular}
\end{table}

\begin{table}[H] 
	\begin{ruledtabular}
	\caption{The distinct symmetries of anisotropic elasticity and crystal systems for 2D materials.}
	\label{tab:2DS}
	\centering
	\tiny 
	\begin{tabular}{c c c c c c c }
		Material  & Crystal  &  {Point } & 2D Space  & No. of independent  & No. of distinct  \\
		symmetry &  system &  group &  group  & elastic constants & Kelvin moduli \\
		\hline Anisotropic & Oblique & C$_1$, C$_2$ & $p1, p2$ & 6 & 3\\
		Orthotropic & Rectangular & D$_1$, D$_2$ & $pm, pg, pmm, pmg$ & 4 & 3\\
		& Centered rectangular & D$_1$, D$_2$ & $cm, cmm, pgg$ &  \\
		Tetragonal  & Square & C$_4$, D$_4$ & $p4, p4m, p4g$ & 3 & 3\\
		Isotropy & Hexagonal & C$_3$, D$_3$, C$_6$, D$_6$ & $p3, p3m1, p31m, p6, p6m$ & 2 & 2\\
	\end{tabular}
	\end{ruledtabular}
\end{table}

The number of independent elastic constants for isotropy only is independent of the orientation of the crystallographic axis system. This is demonstrated in the representation of the stiffness tensor for NiAl, a crystal with cubic symmetry, in \comm{Appendix} \ref{sec:AppA}. Depending on the orientation of the crystal, we have 3 \ref{tab:NiAl1Cij}, 6 \ref{tab:NiAl2Cij} and 9 \ref{tab:NiAl3Cij} distinct elastic constants, respectively. As can be seen, the pattern of the stiffness tensor also changes, but the number of Kelvin moduli remains constant.

\comm{W \cite{Nordmann2018} "Visualising elastic anisotropy: theoretical background and computational implementation", 3 Numerical implementation  bardzo fajnie rozpisane tensory sztywnosci dla roznych symetrii 3D a w \cite{wallace_thermodynamics} Points groups in Tab.1 p.29 
	\cite{kashtalian2013fundamentals} "Fundamentals of the Three-Dimensional Theory of Stability of Deformable Bodies"

W ksiazce \cite{born1988dynamical}"Dynamical Theory of Crystal Lattices",Born, M. and Huang, K. rozwazaja warunki stabilnosci tylko dla symetrii kubicznej

Trudno jednak powiedziec \cite{Zhou1996} "Stability criteria for homogeneously stressed materials and the calculation of elastic constants" 
	
	Praca \cite{Barron1965} "Second-order elastic constants of a solid under stress" moze do zacytowania ale napisana raczej metnie 
	
	W ksiazce \cite{marsden1994mathematical} o tym caly rozdzial 3 
	
	Ladnie opisane w ksiazce \cite{Vannucci2018} "Anisotropic Elasticity, Chapter 2
	General Anisotropic Elasticity"  
	
	Bardzo dobra praca \cite{Wallace1967} "Thermoelasticity of Stressed Materials and Comparison of Various Elastic Constants" 

Zobaczyc \cite{Wang1993} "{Crystal instabilities at finite strain}"- cubic crystal under hydrostatic pressure i \cite{Wang1995} "{Mechanical instabilities of homogeneous crystals}"- we consider a cubic lattice under hydrostatic pressure : Given that B is in general asymmetric, the stability
of B is governed by its symmetrized counterpart,
because
\emph{A}=(1/2)(B$^T$ +B), (2.24)
The stability criterion is then the requirement that all the eigenvalues of \emph{A} be positive

Podobnie:
Under the same conditions define the fully symmetric spatial tangent modulus Bsym:
"Intrinsic elastic stability under controlled Cauchy stress"

To wlasnie chce, bardzo ciekawa praca, stabilnosc dowolnego krysztalu dla roznych deformacji \cite{Morris2000}"The internal stability of an elastic solid", symetryczne B  ... Since 3. has full Voigt symmetry, it can be written as the
6 x 6 matrix A, with eigenvalues A,. Stability is lost when the least of these vanishes
}

\comm{  Moze zrobić jak w "AN ATOMISTIC INSTABILITY CONDITION AND APPLICATIONS"(1)
	Moze to jest dobre? \cite{Wang_2012} "Unifying the criteria of elastic stability of solids"
	
	7.1 Elastic stability conditions of isotropic solids w THEORETICAL STRENGTH OF SOLIDS wang hao 201012 phd 

W \cite{CLAYTON2014104} "{Analysis of intrinsic stability criteria for isotropic third-order Green elastic and compressible neo-Hookean solids}", analizowane 3 intrinsic stability criteria i ta z \cite{Morris2000} i "internal stability
according to strain increments conjugate to Cauchy stress is found to be the most stringent
criterion." 
i podobnie w \cite{Clayton2012} "{Towards a nonlinear elastic representation of finite compression and instability of boron carbide ceramic}",

Baza  the Materials Project (www.materialsproject.org) \cite{Jain2013} sprawdza blednie stabilnosc wg \cite{deJong2015} warunek iii) zapis Voigt'a

Numerycznie \cite{Mota2016} "A Cartesian parametrization for the numerical analysis of material instability" 

Ciekawa praca \cite{Griesser2023} "Analytic elastic coefficients in molecular calculations Finite strain, nonaffine displacements, and many-body interatomic potentials" a zwlaszcza  "H. Stability criteria" First, dynamical or elastic stability are only necessary, not sufficient conditions for stability. This is because stability is governed by H, while regarding either dynamical or elastic stability considers only one of the diagonal blocks H and c
$\Pi$. Those couple through $\Gamma$, and this coupling will generally decrease the lowest individual eigenvalue. 

jesli Elasticity Tensor (clamped ion) to coupling with phonons wiec trzeba uzywac  Elasticity Tensor (relaxed ion) 

\cite{horn13} "{Matrix Analysis}" M is positive definite if and only if all of its eigenvalues are positive. Inne warunki to Cholesky decomposition condition and Sylvester's criterion $https://en.wikipedia.org/wiki/Definite_matrix$

Conditions for positive definiteness and semi-definiteness $https://en.wikipedia.org/wiki/Schur_complement$

- If A is invertible, then X is positive definite if and only if A and its complement X/A are both positive definite:
- If C is invertible, then X is positive definite if and only if C and its complement X/C are both positive definite: 
}
\cite{Vannucci2024} "{Complete Set of Bounds for the Technical Moduli in 3D Anisotropic Elasticity}" : Another possible strategy is to use a mathematical approach: to write the necessary and sufficient conditions for a symmetric matrix to be positive definite. In the most part of texts on continuum mechanics, and not only, this question is linked to a general result on eigenvalues (
being symmetric, the spectral theorem, [3], guarantees the existence of six real eigenvalues): a symmetric matrix is positive definite if and only if all of its eigenvalues are positive. Such theorem finds its deep interest in mechanics, where eigenvalues have specific physical meanings (principal stresses or strains, principal inertia moments, vibration frequencies, critical loads etc.). However, when general conditions are looked for, the analytical expression of the eigenvalues is needed. Unfortunately, the characteristic equation for 
is of degree six and, apart some specially simple cases, its solution is not known.

There is, however, a theorem of linear algebra, much less used in mechanics, known as the Sylvester’s criterion, giving a set of necessary and sufficient conditions for a real symmetric matrix to be positive definite. Quoting from Hohn, [3]: a real quadratic form is positive definite if and only if the leading principal minors of the matrix of the form are all positive.}

\section{Voigt and \emph{orthonormal} notation}
\label{sec:srtn}

In applied mathematics, physics \cite{nye1985physical, Mouhat2014}, continuum and computational mechanics \cite{ting1992anisotropic, Clayton2011}, \emph{ab\,initio} and molecular codes \cite{Plimpton1995}, where symmetric tensors appear, it is common practice to use Voigt notation to reduce their order, so that second-order tensors (strain, stress) are written as vectors and fourth-order tensors (elasticity) as matrices \cite{SINGH2021108068, LIU2022108180}.

\comm{Moze zacytowac \cite{Moakher2008} "{Fourth-order cartesian tensors: old and new facts, notions and applications}" 
	
	Moze sie przydac, ladnie napisane \cite{Man2011} "A Simple Explicit Formula for the Voigt-Reuss-Hill Average of Elastic Polycrystals with Arbitrary Crystal and Texture Symmetries" 

Moze sie przydac opisana symetria 3D i 2D w SzeptynskiP TeoriaSprezystosci.pdf
ladnie opisane 2. A Tensorial Presentation of the Kelvin Formulation w "Adaptation of generalized Hill inequalities to anisotropic elastic symmetries.pdf"

Zacytowano moja prace w \cite{Tang2020} "{Stability, electronic structures, and band alignment of two-dimensional ${\mathrm{II}}_{A}$-IV-${\mathrm{N}}_{2}$ materials}"
	i w "VASPKIT A User-friendly Interface Facilitating High-throughput Computing" Table 3
	In \cite[figure 1]{Haastrup2018} the workflow used to calculate the structure and properties of the materials in C2DB the authors stated that the dynamical stability condition for a structure is not satisfied when elastic constants are negative. Unfortunately, it is an incorrect condition.
	Moreover, in \cite[equation (3)]{Haastrup2018} the authors, for reasons difficult to understand, disregarded shear deformations and calculated only the planar elastic stiffness coefficients $C_{11}$, $C_{22}$, and $C_{12}$, what makes the aforementioned analysis incomplete and insufficient.
	In addition, even these calculated coefficients in C2DB are erroneous, i.e. the stiffness tensor does not have a proper symmetry resulting from the symmetry of the crystal. 
	
	In order to explain what the problem is, some facts from the theory of 2D linear elasticity and elastic stability analysis should be recalled. 

Poprawiono \cite{Haastrup2018} w \cite{Gjerding2021} "{Recent Progress of the Computational 2D Materials Database (C2DB)}" 
}

The most straightforward method for illustrating this notation is to consider the case of a generalized Hooke's law, which describes the linear relationship between strain and stress tensor:
\begin{equation}\label{hl}
	\sigma_{ij}=C_{ijkl}\varepsilon_{kl}\rightarrow \bm{\sigma}=\mathbb{C}\bm{\varepsilon},
\end{equation}
where $\bm{\sigma}$ is the second-order Cauchy stress tensor,
$\mathbb{C}$ is the {fourth-order} anisotropic stiffness tensor and $\bm{\varepsilon}$ is the {second-order} small strain tensor ($i,j,k$=1,2,3 for 3D and $i,j,k$=1,2 for 2D problems), in accordance with the Einstein summation convention, repeated indices are to be understood as implicitly summed.

Since $\bm{\sigma}$ and $\bm{\varepsilon}$ are symmetric tensors, the following holds (minor symmetry)
\begin{equation}\label{cs1}
	C_{ijkl}=C_{jikl}=C_{ijlk},
\end{equation}

\comm{and from the thermodynamic requirement of the existence of a strain energy density function $W(\bm{\varepsilon})$ (hyperelastic material) \cite{Carroll09} such that
\begin{equation}\label{se}
	W=\frac{1}{2}\frac{\partial^2 W}{\partial \varepsilon_{ij} \partial \varepsilon_{kl}}\bigg|_{\bm{\varepsilon}=0}\varepsilon_{ij}\varepsilon_{kl} = \frac{1}{2}C_{ijkl}\varepsilon_{ij}\varepsilon_{kl}\rightarrow W=\frac{1}{2}\bm{\varepsilon}\mathbb{C}\bm{\varepsilon}=\frac{1}{2}\bm{\varepsilon}\bm{\sigma},
\end{equation}
}
{and from the thermodynamic requirement of the existence of a strain energy density function $W(\bm{\varepsilon})$ (hyperelastic material) \cite{Carroll09}} additionally holds (major symmetry)
\begin{equation}\label{cs2}
	C_{ijkl}=C_{klij},
\end{equation}
and hence number of independent components of fourth-order $C_{ijkl}$ reduces to 21 in 3D \cite{hetnarski2010mathematical} and to 6 in 2D \cite{Blinowski1996, He1996}.
\comm{In relations (\ref{hl}) and (\ref{se}) the \emph{orthonormal} notation, employing {fourth-order} Cartesian tensor in three or two dimensions, is used. Also different notations for the generalized Hooke's law, relation (\ref{hl}), are in use.} The non-tensorial Voigt notation mentioned above employs a 6$\times$6 symmetric matrix in 3D:

\begin{equation}
	\centering
	\left[
	\begin{array}{c}
		{\sigma_{11}}\\
		{\sigma_{22}}\\
		{\sigma_{33}}\\
		{\sigma_{23}} \\
		{\sigma_{13}} \\
		{\sigma_{12}} \\
	\end{array}
	\right]=\left[
	\begin{array}{ccc|ccc}
		{C_{1111}} & {C_{1122}} & {C_{1133}} & {C_{1123}} & {C_{1113}} & {C_{1112}}\\
		{C_{1122}} & {C_{2222}} & {C_{2233}} & {C_{2223}} & {C_{2213}} & {C_{2212}}\\
		{C_{1133}} & {C_{2233}} & {C_{3333}} & {C_{3323}} & {C_{3313}} & {C_{3312}}\\ \hline
		{C_{1112}} & {C_{2212}} & {C_{3323}} & {C_{2323}} & {C_{2313}} & {C_{2312}}\\
		{C_{1113}} & {C_{2213}} & {C_{3313}} & {C_{2313}} & {C_{1313}} & {C_{1312}}\\
		{C_{1112}} & {C_{2212}} & {C_{3312}} & {C_{2312}} & {C_{1312}} & {C_{1212}}\\
	\end{array}
	\right] \left[
	\begin{array}{c}
		{\varepsilon_{11}} \\
		{\varepsilon_{22}} \\
		{\varepsilon_{33}} \\
		{2} {\varepsilon_{23}} \\
		{2} {\varepsilon_{13}} \\
		{2} {\varepsilon_{12}} \\
	\end{array}
	\right],\ 
	\label{eqn:Hook3dVoigt}
\end{equation}

or
\begin{equation}
	\centering
	\left[
	\begin{array}{c}
		{\hat\sigma_{1}}\\
		{\hat\sigma_{2}}\\
		{\hat\sigma_{3}}\\
		{\hat\sigma_{4}} \\
		{\hat\sigma_{5}} \\
		{\hat\sigma_{6}} \\
	\end{array}
	\right]=\left[
	\begin{array}{cccccc}
		{\hat{C}_{11}} & {\hat{C}_{12}} & {\hat{C}_{13}} & {\hat{C}_{14}} & {\hat{C}_{15}} & {\hat{C}_{16}}\\
		{\hat{C}_{12}} & {\hat{C}_{22}} & {\hat{C}_{23}} & {\hat{C}_{24}} & {\hat{C}_{25}} & {\hat{C}_{26}}\\
		{\hat{C}_{13}} & {\hat{C}_{23}} & {\hat{C}_{33}} & {\hat{C}_{34}} & {\hat{C}_{35}} & {\hat{C}_{36}}\\ 
		{\hat{C}_{14}} & {\hat{C}_{24}} & {\hat{C}_{34}} & {\hat{C}_{44}} & {\hat{C}_{45}} & {\hat{C}_{46}}\\
		{\hat{C}_{15}} & {\hat{C}_{25}} & {\hat{C}_{35}} & {\hat{C}_{45}} & {\hat{C}_{55}} & {\hat{C}_{56}}\\
		{\hat{C}_{16}} & {\hat{C}_{26}} & {\hat{C}_{36}} & {\hat{C}_{46}} & {\hat{C}_{56}} & {\hat{C}_{66}}\\
	\end{array}
	\right] \left[
	\begin{array}{c}
		{\hat\varepsilon_{1}} \\
		{\hat\varepsilon_{2}} \\
		{\hat\varepsilon_{3}} \\
		{\hat\gamma_{4}} \\
		{\hat\gamma_{5}} \\
		{\hat\gamma_{6}} \\
	\end{array}
	\right]\rightarrow \bm{\hat{\sigma}}=\bm{\hat{C}}\bm{\hat{\varepsilon}}.\  
	\label{eqn:Hook3dVoigt2}
\end{equation}

In 2D 3$\times$3 symmetric matrix:
\begin{equation}
	\centering
	\left[
	\begin{array}{c}
		{\sigma_{11}}\\
		{\sigma_{22}}\\
		{\sigma_{12}} \\
	\end{array}
	\right]=\left[
	\begin{array}{cc|c}
		{C_{1111}} & {C_{1122}} & {C_{1112}} \\
		{C_{1122}} & {C_{2222}} & {C_{2212}} \\\hline
		{C_{1112}} & {C_{2212}} & {C_{1212}} \\
	\end{array}
	\right] \left[
	\begin{array}{c}
		{\varepsilon_{11}} \\
		{\varepsilon_{22}} \\
		2{\varepsilon_{12}} \\
	\end{array}
	\right],\
	\label{eqn:HookVoigt}
\end{equation}

or
\begin{equation}
	\centering
	\left[
	\begin{array}{c}
		{\hat{\sigma}_{1}}\\
		{\hat{\sigma}_{2}}\\
		{\hat{\sigma}_{3}} \\
	\end{array}
	\right]=\left[
	\begin{array}{ccc}
		{\hat{C}_{11}} & {\hat{C}_{12}} & {\hat{C}_{13}} \\
		{\hat{C}_{12}} & {\hat{C}_{22}} & {\hat{C}_{23}} \\
		{\hat{C}_{13}} & {\hat{C}_{23}} & {\hat{C}_{33}} \\
	\end{array}
	\right] \left[
	\begin{array}{c}
		{\hat{\varepsilon}_{1}} \\
		{\hat{\varepsilon}_{2}} \\
		{\hat{\gamma}_{3}} \\
	\end{array}
	\right]\rightarrow \bm{\hat{\sigma}}=\bm{\hat{C}}\bm{\hat{\varepsilon}}.\ 
	\label{eqn:HookVoigt2}
\end{equation}

In this notation, the non-diagonal elements of the second-order strain tensor are doubled and denoted as $\hat{\gamma}_{J}$, and have the interpretation of engineering shear strains, for stresses there is no doubling.

The less popular is an \emph{orthonormal}, also called Mandel, notation in 3D:

\begin{equation}
	\centering
	\left[
	\begin{array}{c}
		{\sigma_{11}}\\
		{\sigma_{22}}\\
		{\sigma_{33}}\\
		\sqrt{2} {\sigma_{23}} \\
		\sqrt{2} {\sigma_{13}} \\
		\sqrt{2} {\sigma_{12}} \\
	\end{array}
	\right]=\left[
	\begin{array}{ccc|ccc}
		{C_{1111}} & {C_{1122}} & {C_{1133}} & \sqrt{2}{C_{1123}} & \sqrt{2}{C_{1113}} & \sqrt{2}{C_{1112}}\\
		{C_{1122}} & {C_{2222}} & {C_{2233}} & \sqrt{2}{C_{2223}} & \sqrt{2}{C_{2213}} & \sqrt{2}{C_{2212}}\\
		{C_{1133}} & {C_{2233}} & {C_{3333}} & \sqrt{2}{C_{3323}} & \sqrt{2}{C_{3313}} & \sqrt{2}{C_{3312}}\\ \hline
		\sqrt{2}{C_{1112}} & \sqrt{2}{C_{2212}} & \sqrt{2}{C_{3323}} & 2{C_{2323}} &  2{C_{2313}} &  2{C_{2312}}\\
		\sqrt{2}{C_{1113}} & \sqrt{2}{C_{2213}} & \sqrt{2}{C_{3313}} & 2{C_{2313}} & 2{C_{1313}} & 2{C_{1312}}\\
		\sqrt{2}{C_{1112}} & \sqrt{2}{C_{2212}} & \sqrt{2}{C_{3312}} & 2{C_{2312}} & 2{C_{1312}} & 2{C_{1212}}\\
	\end{array}
	\right] \left[
	\begin{array}{c}
		{\varepsilon_{11}} \\
		{\varepsilon_{22}} \\
		{\varepsilon_{33}} \\
		\sqrt{2} {\varepsilon_{23}} \\
		\sqrt{2} {\varepsilon_{13}} \\
		\sqrt{2} {\varepsilon_{12}} \\
	\end{array}
	\right],\ 
	\label{eqn:Hook3dOrthonormal}
\end{equation}

or
\begin{equation}
	\centering
	\left[
	\begin{array}{c}
		\sigma_{1}\\
		\sigma_{2}\\
		\sigma_{3}\\
		\sigma_{4}\\
		\sigma_{5}\\
		\sigma_{6}\\
	\end{array}
	\right]=\left[
	\begin{array}{cccccc}
		{C_{11}} & {C_{12}} & {C_{13}} & {C_{14}} & {C_{15}} & {C_{16}}\\
		{C_{12}} & {C_{22}} & {C_{23}} & {C_{24}} & {C_{25}} & {C_{26}}\\
		{C_{13}} & {C_{23}} & {C_{33}} & {C_{34}} & {C_{35}} & {C_{36}}\\
		{C_{14}} & {C_{24}} & {C_{34}} & {C_{44}} & {C_{45}} & {C_{46}}\\
		{C_{15}} & {C_{25}} & {C_{35}} & {C_{45}} & {C_{55}} & {C_{56}}\\
		{C_{16}} & {C_{26}} & {C_{36}} & {C_{46}} & {C_{56}} & {C_{66}}\\
	\end{array}
	\right] \left[
	\begin{array}{c}
		{\varepsilon_{1}} \\
		{\varepsilon_{2}} \\
		{\varepsilon_{3}} \\
		{\varepsilon_{4}} \\
		{\varepsilon_{5}} \\
		{\varepsilon_{6}} \\
	\end{array}
	\right]\rightarrow \bm{\tilde{\sigma}}=\bm{\tilde{C}}\bm{\tilde{\varepsilon}}.\  
	\label{eqn:Hook3dOrthonormal2}
\end{equation}

In 2D:
\begin{equation}
	\centering
	\left[
	\begin{array}{c}
		{\sigma_{11}}\\
		{\sigma_{22}}\\
		\sqrt{2} {\sigma_{12}} \\
	\end{array}
	\right]=\left[
	\begin{array}{cc|c}
		{C_{1111}} & {C_{1122}} & \sqrt{2}{C_{1112}} \\
		{C_{1122}} & {C_{2222}} & \sqrt{2}{C_{2212}} \\ \hline
		\sqrt{2}{C_{1112}} & \sqrt{2}{C_{2212}} & 2{C_{1212}} \\
	\end{array}
	\right] \left[
	\begin{array}{c}
		{\varepsilon_{11}} \\
		{\varepsilon_{22}} \\
		\sqrt{2} {\varepsilon_{12}} \\
	\end{array}
	\right],\ 
	\label{eqn:HookOrthonormal}
\end{equation}

or
\begin{equation}
	\centering
	\left[
	\begin{array}{c}
		{{\sigma}_{1}}\\
		{{\sigma}_{2}}\\
		{{\sigma}_{3}} \\
	\end{array}
	\right]=\left[
	\begin{array}{ccc}
		{{C}_{11}} & {{C}_{12}} & {{C}_{13}} \\
		{{C}_{12}} & {{C}_{22}} & {{C}_{23}} \\ 
		{{C}_{13}} & {{C}_{23}} & {{C}_{33}} \\
	\end{array}
	\right] \left[
	\begin{array}{c}
		{{\varepsilon}_{1}} \\
		{{\varepsilon}_{2}} \\
		{{\varepsilon}_{3}} \\
	\end{array}
	\right]\rightarrow \bm{\tilde{\sigma}}=\bm{\tilde{C}}\bm{\tilde{\varepsilon}}.\ 
	\label{eqn:HookOrthonormal2}
\end{equation}

The difference between the Voigt notation and the \emph{orthonormal} notation is not only the factor of 2 and its square root, but it is much more fundamental.
In the {Voigt} notation, the elements of the matrix $\bm{\hat{C}}$ in Eqs.(\ref{eqn:Hook3dVoigt2}) and (\ref{eqn:HookVoigt2}) are not the elements of a {2nd-order} tensor, while in the \emph{orthonormal} notation, the elements of $\bm{\tilde{C}}$ in Eqs.(\ref{eqn:Hook3dOrthonormal2}) and (\ref{eqn:HookOrthonormal2}) are the elements of a {2nd-order tensor} in six dimensions for 3D and three dimensions for 2D problems. The {fourth-order} tensor notation (\ref{hl}) and \emph{orthonormal} notation (\ref{eqn:Hook3dOrthonormal2}, \ref{eqn:HookOrthonormal2}) are tensorially equivalent \cite{MEHRABADI1990, Blinowski1996}. {It is also worth mentioning that in this notation the stiffness tensor $\bm{\tilde{C}}$ and the compliance tensor $\bm{\tilde{S}}$ have the same bases, are collinear (have the same eigenvectors and the eigenvalues of $\bm{\tilde{S}}$ are equal to the inverse of the eigenvalues of $\bm{\tilde{C}}$).} There is also an equivalence of quadratic norms here \cite{Morin2020}:

\comm{W pracy \cite{Morin2020} "Generalized Euclidean Distances for Elasticity Tensors" wzory 18-24, fajnie zapisana rownowaznosc norm dla {fourth-order} and Kelvin notations
Podobnie w "Two-dimensional Hooke’s tensors—Isotropic decomposition, effective symmetry criteria" 3.15 do 3.21}

\begin{equation}\label{s3}
	\|\bm{\sigma}\|=\sqrt{\sigma_{ij}\sigma_{ij}}= \|\bm{\tilde{\sigma}}\|=\sqrt{{\sigma}_{I}{\sigma}_{I}} ,
\end{equation}

\begin{equation}\label{e3}
	\|\bm{\varepsilon}\|=\sqrt{\varepsilon_{ij}\varepsilon_{ij}}= \|\bm{\tilde{\varepsilon}}\|=\sqrt{{\varepsilon}_{I}{\varepsilon}_{I}} ,
\end{equation}

\begin{eqnarray}\label{c6}
	\|\mathbb{C}\|=\sqrt{C_{ijkl}C_{ijkl}}= \|\bm{\tilde{C}}\| =\sqrt{{C}_{IJ}{C}_{IJ}} .
\end{eqnarray}

For Voigt notation such equivalence of norms does not occur, see the Supplementary material \ref{sec:ss} \comm{\href{run:./Mechanical stability.nb}{Supplementary Material}}.


\section{Internal stability criterion for stress-free lattice}
\label{sec:Isc}

The case of mechanical stability conditions for an unstressed system is considered first. These stability conditions are contained here in the requirement that the local quadratic approximation of the strain energy density function, $W(\bm{\varepsilon})$, is always positive, strictly convex:

\begin{equation}\label{se2}
	W({\varepsilon_{ij}}) = \frac{1}{2}\frac{\partial^2 W}{\partial \varepsilon_{ij} \partial \varepsilon_{kl}}\bigg|_{\bm{\varepsilon}=0}\varepsilon_{ij}\varepsilon_{kl} = \frac{1}{2}C_{ijkl}\varepsilon_{ij}\varepsilon_{kl}\rightarrow W(\bm{\varepsilon})=\frac{1}{2}\bm{\varepsilon}\mathbb{C}\bm{\varepsilon}=\frac{1}{2}\bm{\varepsilon}\bm{\sigma}.
\end{equation}
The meaning and properties of the quantities used here have already been explained in Sec.~\ref{sec:srtn}. However, this fourth-order tensor condition cannot be used directly. It is typically transformed into a matrix condition using Voigt notation. Then the positivity of the stiffness matrix, which is not a tensor in Voigt notation, is checked using the condition for positivity of the leading principal minors \cite{Mouhat2014, ting1992anisotropic, Gao2023}. This condition is known as Sylvester's criterion and is equivalent to checking whether an upper triangular matrix has all positive diagonal elements after LU decomposition \cite{horn13}. 
Still, there are several problems with this approach, it is not tensorially equivalent to the original problem, it depends on the choice of the base of the lattice vectors, and for low symmetries it is cumbersome. This will be clarified after discussing an alternative approach.  

For an even-order tensor, the eigenvalue problem is well posed \cite{RYCHLEWSKI1984303, KO2009note}, and the spectral decomposition of stiffness tensors can be equivalently expressed in both the fourth-order \cite{Moakher2006} and \emph{orthonormal} notation: 


\begin{equation}\label{eig34}
	\mathbb{C} = \sum_{I=1}^{6(3)} \lambda_I \bm{\varepsilon_I}\otimes\bm{\varepsilon_I} \bm{\leftrightarrow} \bm{\tilde{C}} = \sum_{I=1}^{6(3)} \lambda_I \bm{\tilde{\varepsilon_I}}\otimes\bm{\tilde{\varepsilon_I}},
\end{equation} 

where $\lambda_I$ are called the stiffness, Kelvin moduli \cite{Thomson1856} of $\mathbb{C}$ or $\bm{\tilde{C}}$, a $\bm{\varepsilon_I}$ here are the second-order symmetric eigenstate, eigenstrain tensors of $\mathbb{C}$, and $\bm{\tilde{\varepsilon_I}}$ are eigenstate, eigenstrain vectors of $\bm{\tilde{C}}$ written in \emph{orthonormal} notation ($I$=1,\dots,6 for 3D (Eqs.\ref{eqn:Hook3dOrthonormal} and \ref{eqn:Hook3dOrthonormal2}) and $I$=1,\dots,3 for 2D (Eqs.\ref{eqn:HookOrthonormal} and \ref{eqn:HookOrthonormal2})). The condition of the positivity of the Kelvin moduli is an alternative to the Sylvester criterion and guarantees the required positivity of the quadratic form in Eq.\ref{se2}. And this approach, based on the spectral decomposition of the fourth-order stiffness tensors mapped to second-order tensors using \emph{orthonormal} notation, and the verification of the positivity of the so-called Kelvin moduli, will be dominant in this study. 
It is worth making two further points before moving on to other considerations. First, the previously mentioned condition of positive slopes of acoustic branches of phonons at $\Gamma$ point corresponds to strong ellipticity in mathematical elasticity. This condition does not imply positivity of the strain energy density function, $W(\bm{\varepsilon})$ in Eq.\ref{se2}, but the opposite implication appears \cite{hetnarski2010mathematical}. The strong ellipticity condition allows the existence of a negative bulk modulus $B$ and Poisson's ratio $\nu$, which is only limited by the condition $\notin [\frac{1}{2},1]$, which seems to be unphysical \cite{Ogden1997}. Second, only the case of relaxed ions in the computational cell is considered here, i.e., when internal atomic coordinates are relaxed. In the case of clamped ions, without internal relaxation of the atoms, the mechanical and dynamical stability cannot be decoupled and the mechanical-phonon coupled system must be considered \cite{Griesser2023}.\\
The symmetries of the stiffness tensor will now be discussed. Aspects of symmetry are known to be important in studying physical phenomena \cite{nye1985physical}. Crystal point group symmetry determines symmetry of physical properties. The principle of symmetry (Neumann’s Principle) directly states that: "The symmetry elements of any macroscopic physical property of a crystal must include the
symmetry elements of the point group of the crystal" \cite{malgrange2014symmetry}. Since physical properties are represented by tensors, tensors for crystals must also have corresponding symmetries \cite{Newnham2004}.
However, the symmetry classification of linear elastic materials is not related to crystallography. This is due to the properties of {fourth-order} Euclidean symmetric tensors (from the linearity of phenomenological Hooke's law and the properties of two, three-dimensional Euclidean space)\cite{Mazdziarz15}. For 3D linear hyperelastic materials, there are eight classes of symmetry \cite{CHADWICK20012471} and four classes of symmetry for 2D \cite{Blinowski1996, Mazdziarz2019}. A classification of all symmetries of anisotropic elasticity, their relations to crystal systems, point and space groups, the corresponding number of distinct elastic constants and Kelvin moduli is gathered in Table \ref{tab:3DS} for 3D materials and Table \ref{tab:2DS} for 2D materials.

\comm{Problemy z warunkami z \cite{Mouhat2014, ting1992anisotropic, Gao2023} : Voigt notation, bazuja na the leading principal minors of C complexity is $O$(n$^3$) , warunki sa wazne tylko dla jednej bazy wektorow sieci nie sa ogolne , dotyczy to tez warunkow dla symetrii kubicznej z \cite{Born1940}

 \cite{ELLIOTT2006161} "{Stability of crystalline solids—I: Continuum and atomic lattice considerations}" There is overlap between the phonon-stability and CB-stability criteria. If one
	considers perturbations with no uniform deformation component, dU  0, then the components q2~E=qs2 of the CB-stability matrix, Eq. (2.64) (in the limit as n!1, for the infinite crystal), are positive definite only if the optic phonon equations (2.53) have positive eigenvalues, . 
	The phonon (dynamical)- and mechanical (Born)-stability criteria are, in a sense, complementary. There are 3 acoustic branches of the phonon dispersion relation that start linearly at $\Gamma$ point. The slope of the dispersion curve near $\Gamma$ is the speed of sound. Knowing these slopes, the point symmetry of the crystal and its density, we can deduce the full stiffness tensor. The dynamical stability condition requires that all phonon modes have positive frequencies. Positive slopes of acoustic branches at $\Gamma$ point do not imply mechanical stability \cite{Cowley1976}, these conditions are more complicated..... Part II "the
	CB-stability criterion and the phonon-stability criterion must be satisfied. This ensures
	stability with respect to all quasi-uniform perturbations (CB) (3.7) and bounded displacement
	perturbations of all wavelengths (phonon)"

	$https://www.ctcms.nist.gov/potentials/atomman/tutorial/3.1._ElasticConstants_class.html$
	
	Uzywaja powszechnie {Voigt} notation \cite{Sinko2002}-(stabilnosc tylko pod cisnieniem) 
	
	\cite{Mouhat2014} \cite{ting1992anisotropic}-the condition for positivity of all the principal minors
	
	
	\cite{Gao2023} "Elastic stability criteria of seven crystal systems and their application under pressure: Taking carbon as an example", Void notation, for low-symmetry systems warunki sa cumbersome i wyprowadzenie jest prawdziwe tylko okreslonej reprezentacji Cij czyli only for a specific orientation of the crystallographic axes ... "all the leading principal minors of C (determinants of its upper-left k by k sub-matrix, 1 $\leq$ k $\leq$ 6) are positive." , complexity is $O$(n$^3$)

	\cite{SINGH2021108068} "MechElastic: A Python Library for Analysis of Mechanical and Elastic Properties of Bulk and 2D Materials" - cytuje mnie, Voigt notation
	\cite{LIU2022108180} "ElasTool: An automated toolkit for elastic constants calculation" -> Voigt notation

    Fajna praca \cite{Chadwick1971} "{On the definition of elastic moduli}"

and similarly under hydrostatic pressure in \cite{Grimvall2012}

"Lattice instabilities in metallic elements" Opis wykorzystac: Here Cijkl are elements in the elastic constant tensor C evaluated at the current stressed state (which may not have cubic symmetry), $\sigma_{ij}$ specify the external stresses

\cite{CHADWICK20012471} "A new proof that the number of linear elastic symmetries is eight"

Prawie wszystko co mi potrzeba dla 3D \cite{Cowin1995} "Anisotropic symmetries of linear elasticity" wzor (15)

Moze przydatne \cite{Bona2007} "Coordinate-free Characterization of the Symmetry Classes of Elasticity Tensors": Fig. 1 Diagram representing relations among the symmetry classes


Ciekawe \cite{Bigoni2000} "Bifurcation and Instability of Non-Associative Elastoplastic Solids" str 25 (93) pokazano ze positive definiteness of C (PD condition) implies strong ellipticity (SE) 	

Warto moze douczyc sie Harmonic and Cartan decompositions \cite{Forte1996} "{Symmetry classes for elasticity tensors}" i \cite{Wen2024} "{An equivariant graph neural network for the elasticity tensors of all seven crystal systems}"
https://github.com/wengroup/matten?tab=readme-ov-file
	
	Sprawdzic co naprawde jest liczone w ABINIT i porownac to z elasticity-oganov.pdf i  https://docs.abinit.org/tutorial/elastic/index.html 
	
	Autorzy \cite{Nielsen1983} "{First-Principles Calculation of Stress}" licza virial stress, ktory odpowiada Cauchy stress Zdaja sobie sprawe z istnienia skonczonych deformacji, ale przeliczaja male na duze tylko dla jednowymiarowego przypadku
	
	Fajne \cite{FEA2017} "Nonlinear finite elements$/$Rate form of hyperelastic laws" 
	
	W \cite{Nagel2016} "On advantages of the Kelvin mapping in finite element implementations of deformation processes"  ciekawa Table 1 Comparison of different tensor norms

	Dla 3D przepisac z \cite{BoehlkeBrueggemann2001} "Graphical Representation of the Generalized Hooke's Law", przepisac (8), (9)-(16)
	
	Przydatne \cite{KO2009note} i "Waves in Nonlinear Pre-Stressed Materials" rozdział I

	pomysl jak u mnie tylko Voigt notation and uniaxial stress \cite{Wen2017} "{Lattice stability and the effect of Co and Re on the ideal strength of Ni: First-principles study of uniaxial tensile deformation}": To test the mechanical stability of a lattice during the tensile processes, we calculate the elastic constants Cij of the lattice at every tensile strain and examine the mechanical stability
	conditions for the corresponding crystal according to Eqs. (9), (13), and (15). ... czy wzor (4) jak u mnie? 
}

\begin{table}[H] 
		\caption{The distinct symmetries of anisotropic elasticity and crystal systems for 3D materials.}
		\label{tab:3DS}
		\centering
		\setlength{\tabcolsep}{4.0pt} 
		\tiny 
		\begin{tabular}{c c c c c c }
			\hline
			Material  & Crystal  &  {Point } & Space  & No. of independent & No. of distinct  \\
			symmetry &  system &  group &  group & elastic constants & Kelvin moduli \\
			\hline Triclinic & Triclinic & C$_1$, C$_i$ & 1-2 & 21 & 6 \\
			Monoclinic & Monoclinic & C$_2$, C$_s$, C$_{2h}$ & 3–15 & 13 & 6 \\
			Orthotropic & Orthorhombic & D$_2$, C$_{2v}$, D$_{2h}$ & 16-74 & 9 & 6\\
			Tetragonal & Tetragonal & C$_4$, S$_4$, C$_{4h}$, D$_4$ & 75–142 & 6 & 5\\
			&  & C$_{4v}$, D$_{2d}$, D$_{4h}$ & &  \\
			Trigonal & Trigonal & C$_3$, S$_6$, D$_3$, C$_{3v}$, D$_{3d}$ & 143–167 & 6 & 4\\
			Transverse Isotropy & Hexagonal & C$_6$, C$_{3h}$, C$_{6h}$, D$_6$ & 168–194 & 5 & 4\\
			&  & C$_{6v}$, D$_{3h}$, D$_{6h}$ & &  \\
			Cubic & Cubic & T, T$_h$, O, T$_d$, O$_h$ & 195-230 & 3 & 3\\
			Isotropy & & &  & 2 & 2\\
		\hline 
		\end{tabular}
\end{table}

\begin{table}[H] 
		\caption{The distinct symmetries of anisotropic elasticity and crystal systems for 2D materials.}
		\label{tab:2DS}
		\centering
		\setlength{\tabcolsep}{2.0pt} 
		\tiny 
		\begin{tabular}{c c c c c c c }
			\hline
			Material  & Crystal  &  {Point } & 2D Space  & No. of independent  & No. of distinct  \\
			symmetry &  system &  group &  group  & elastic constants & Kelvin moduli \\
			\hline Anisotropic & Oblique & C$_1$, C$_2$ & $p1, p2$ & 6 & 3\\
			Orthotropic & Rectangular & D$_1$, D$_2$ & $pm, pg, pmm, pmg$ & 4 & 3\\
			& Centered rectangular & D$_1$, D$_2$ & $cm, cmm, pgg$ &  \\
			Tetragonal  & Square & C$_4$, D$_4$ & $p4, p4m, p4g$ & 3 & 3\\
			Isotropy & Hexagonal & C$_3$, D$_3$, C$_6$, D$_6$ & $p3, p3m1, p31m, p6, p6m$ & 2 & 2\\
		\hline
		\end{tabular}
\end{table}

The number of distinct elastic constants requires some comment. First, it is independent of the orientation of the crystallographic axis system only for isotropy.
This is demonstrated in the representation of the stiffness tensor for NiAl, a crystal with cubic symmetry, in \comm{Appendix} \ref{sec:AppA}. Depending on the orientation of the crystal, we have 3 \ref{tab:NiAl1Cij}, 6 \ref{tab:NiAl2Cij} and 9 \ref{tab:NiAl3Cij} distinct elastic constants, respectively. As can be seen, the pattern of the stiffness tensor also changes, but the number of Kelvin moduli remains constant.
In addition to rotating the structure in molecular or \emph{ab\,initio} calculations, we don't always use conventional computational cells. Sometimes it is more convenient to convert non-orthogonal cells, such as hexagonal or trigonal cells, into orthogonal cells \cite{MAZDZIARZ2024109953}.
Second, some authors introduce two separate tetragonal and two separate trigonal symmetries, differing in the number of distinct elastic constants \cite{nye1985physical, Mouhat2014, ZHANG2017403}. Such a separation is unnecessary. They all have 6 distinct elastic constants when the proper substitution is made \cite{Cowin1995}.   


\comm{W \cite{Nordmann2018} "Visualising elastic anisotropy: theoretical background and computational implementation", 3 Numerical implementation  bardzo fajnie rozpisane tensory sztywnosci dla roznych symetrii 3D a w \cite{wallace_thermodynamics} Points groups in Tab.1 p.29 
	\cite{kashtalian2013fundamentals} "Fundamentals of the Three-Dimensional Theory of Stability of Deformable Bodies"
	
	W ksiazce \cite{born1988dynamical}"Dynamical Theory of Crystal Lattices",Born, M. and Huang, K. rozwazaja warunki stabilnosci tylko dla symetrii kubicznej

	Trudno jednak powiedziec \cite{Zhou1996} "Stability criteria for homogeneously stressed materials and the calculation of elastic constants" 
	
	Praca \cite{Barron1965} "Second-order elastic constants of a solid under stress" moze do zacytowania ale napisana raczej metnie 
	
	W ksiazce \cite{marsden1994mathematical} o tym caly rozdzial 3 
	
	Ladnie opisane w ksiazce \cite{Vannucci2018} "Anisotropic Elasticity, Chapter 2
	General Anisotropic Elasticity"  
	
	Bardzo dobra praca \cite{Wallace1967} "Thermoelasticity of Stressed Materials and Comparison of Various Elastic Constants"

Dla 3D "Table 3. Classification of crystal systems, point-group symbols and space-group number (S.G.N) are provided with the number of independent elastic constants for 3D materials. The material prototypes are shown in the last column." i 2D "Table 4. Classification of crystal systems, 2D space groups and the independent elastic constants are provided with the number of independent elastic constants for 2D materials. The material prototypes are shown in the last column." w AELAS \cite{ZHANG2017403} (Table 5 lattice vectors for 3D materials i Table 6 lattice vectors for 2D materials) Podzial na trig I i trig II, oraz tetra I i tetra II jest bledny \cite{Cowin1995}, wszystkie maja 6 stałych sprezystosci trzeba tylko dokonac wlasciwego podstawienia, tu tez blad \cite{Mouhat2014} It is also shown that the restrictions on the elastic coefficients appearing in Hooke's
law follow in a simple and straightforward fashion from orthogonal transformations based on a
small subset of the small catalogue of planes of mirror symmetry-Cowin

Zobaczyc \cite{Wang1993} "{Crystal instabilities at finite strain}"- cubic crystal under hydrostatic pressure i \cite{Wang1995} "{Mechanical instabilities of homogeneous crystals}"- we consider a cubic lattice under hydrostatic pressure : Given that B is in general asymmetric, the stability
of B is governed by its symmetrized counterpart,
because
\emph{A}=(1/2)(B$^T$ +B), (2.24)
The stability criterion is then the requirement that all the eigenvalues of \emph{A} be positive

Podobnie:
Under the same conditions define the fully symmetric spatial tangent modulus Bsym:
"Intrinsic elastic stability under controlled Cauchy stress"

To wlasnie chce, bardzo ciekawa praca, stabilnosc dowolnego krysztalu dla roznych deformacji \cite{Morris2000}"The internal stability of an elastic solid", symetryczne B  ... Since 3. has full Voigt symmetry, it can be written as the
6 x 6 matrix A, with eigenvalues A,. Stability is lost when the least of these vanishes

	 Moze zrobić jak w "AN ATOMISTIC INSTABILITY CONDITION AND APPLICATIONS"(1)
	Moze to jest dobre? \cite{Wang_2012} "Unifying the criteria of elastic stability of solids"
	
	7.1 Elastic stability conditions of isotropic solids w THEORETICAL STRENGTH OF SOLIDS wang hao 201012 phd 
	
	W \cite{CLAYTON2014104} "{Analysis of intrinsic stability criteria for isotropic third-order Green elastic and compressible neo-Hookean solids}", analizowane 3 intrinsic stability criteria i ta z \cite{Morris2000} i "internal stability
	according to strain increments conjugate to Cauchy stress is found to be the most stringent
	criterion." 
	i podobnie w \cite{Clayton2012} "{Towards a nonlinear elastic representation of finite compression and instability of boron carbide ceramic}",

Baza  the Materials Project (www.materialsproject.org) \cite{Jain2013} sprawdza blednie stabilnosc wg \cite{deJong2015} warunek iii) zapis Voigt'a

Numerycznie \cite{Mota2016} "A Cartesian parametrization for the numerical analysis of material instability" 

Ciekawa praca \cite{Griesser2023} "Analytic elastic coefficients in molecular calculations Finite strain, nonaffine displacements, and many-body interatomic potentials" a zwlaszcza  "H. Stability criteria" First, dynamical or elastic stability are only necessary, not sufficient conditions for stability. This is because stability is governed by H, while regarding either dynamical or elastic stability considers only one of the diagonal blocks H and c
$\Pi$. Those couple through $\Gamma$, and this coupling will generally decrease the lowest individual eigenvalue. 

jesli Elasticity Tensor (clamped ion) to coupling with phonons wiec trzeba uzywac  Elasticity Tensor (relaxed ion) 

\cite{horn13} "{Matrix Analysis}" M is positive definite if and only if all of its eigenvalues are positive. Inne warunki to Cholesky decomposition condition and Sylvester's criterion $https://en.wikipedia.org/wiki/Definite_matrix$

Conditions for positive definiteness and semi-definiteness $https://en.wikipedia.org/wiki/Schur_complement$

- If A is invertible, then X is positive definite if and only if A and its complement X/A are both positive definite:
- If C is invertible, then X is positive definite if and only if C and its complement X/C are both positive definite: 

\cite{Vannucci2024} "{Complete Set of Bounds for the Technical Moduli in 3D Anisotropic Elasticity}" : Another possible strategy is to use a mathematical approach: to write the necessary and sufficient conditions for a symmetric matrix to be positive definite. In the most part of texts on continuum mechanics, and not only, this question is linked to a general result on eigenvalues (
being symmetric, the spectral theorem, [3], guarantees the existence of six real eigenvalues): a symmetric matrix is positive definite if and only if all of its eigenvalues are positive. Such theorem finds its deep interest in mechanics, where eigenvalues have specific physical meanings (principal stresses or strains, principal inertia moments, vibration frequencies, critical loads etc.). However, when general conditions are looked for, the analytical expression of the eigenvalues is needed. Unfortunately, the characteristic equation for 
is of degree six and, apart some specially simple cases, its solution is not known.

There is, however, a theorem of linear algebra, much less used in mechanics, known as the Sylvester’s criterion, giving a set of necessary and sufficient conditions for a real symmetric matrix to be positive definite. Quoting from Hohn, [3]: a real quadratic form is positive definite if and only if the leading principal minors of the matrix of the form are all positive.


}
\comm
{\section{3D}
\label{sec:3D}

The following representations of stiffness tensor are given with respect to the symmetry axes, in a canonical base, with standard lattice vectors \cite{ZHANG2017403}. In molecular or \emph{ab\,initio} calculations, we don't always use conventional computational cells. Sometimes it is more convenient to convert non-orthogonal cells, such as hexagonal or trigonal cells, into orthogonal cells \cite{MAZDZIARZ2024109953}. 



\begin{enumerate}[I.]
	
	\item Cubic \& Isotropy $\rightarrow$ Cubic lattice \newline  
	(3 \& 2  elastic constants; 3 \&2 Kelvin moduli)
\begin{equation}
		\centering
		\left[\bm{\tilde{C}}_{\alpha\beta}\right]\rightarrow\left[
		\begin{array}{cccccc}
			{C_{11}}  & {C_{12}} & {C_{12}} & {0}      & {0}      & {0} \\
			{C_{12}}  & {C_{11}} & {C_{12}} & {0}      & {0}      & {0} \\
			{C_{12}}  & {C_{12}} & {C_{11}} & {0}      & {0}      & {0} \\
			{0}       & {0}      & {0}      & {C_{44}} & {0}      & {0}    \\
			{0}       & {0}      & {0}      & {0}      & {C_{44}} & {0}    \\
			{0}       & {0}      & {0}      & {0}      & {0}      & {C_{44}}\\
		\end{array}
		\right]\,.
		\label{tab:CubicCij}
\end{equation}
	
	(For Isotropy $C_{44}=C_{11}-C_{12}$)
	
	$C_{11}-C_{12}>0$ \& $C_{11}+2C_{12}>0$ \& $C_{44}>0$ or\\
	$\lambda_{I, II, III}=C_{44}>0$ \& $\lambda_{IV, V}=(C_{11}-C_{12})>0$ \& $\lambda_{VI}=(C_{11}+2C_{12})>0$.

	\item Transverse Isotropy $\rightarrow$ Hexagonal lattice \newline  
	(5 elastic constants; 4 Kelvin moduli)
\begin{equation}
		\centering
		\left[\bm{\tilde{C}}_{\alpha\beta}\right]\rightarrow\left[
		\begin{array}{cccccc}
			{C_{11}} & {C_{12}} & {C_{13}} & {0} & {0} & {0} \\
			{C_{12}} & {C_{11}} & {C_{13}} & {0} & {0} & {0} \\
			{C_{13}} & {C_{13}} & {C_{33}} & {0} & {0} & {0} \\
			{0} & {0} & {0} & {C_{44}} & {0} & {0} \\
			{0} & {0} & {0} & {0} & {C_{44}} & {0} \\
			{0} & {0} & {0} & {0} & {0} & {C_{11}-C_{12}} \\
		\end{array}
		\right]\,.
		\label{tab:TICij}
\end{equation}
	
	$C_{11} > \left|C_{12}\right|$ \& $2C_{13}^2>C_{33}(C_{11}+C_{12})$ \& $C_{44}>0$ \& $(C_{11}-C_{12})>0$ or\\
	$\lambda_{I, II}=C_{44}>0$ \& $\lambda_{III, IV}=(C_{11}-C_{12})>0$ \& \\ 
	$\lambda_{V}=\frac{1}{2}\left(C_{11}+C_{22}+C_{33}-\bigstar\right)>0$ \& 
	$\lambda_{VI}=\frac{1}{2}\left(C_{11}+C_{22}+C_{33}+\bigstar\right)>0$, where $\bigstar=\sqrt{C_{11}^2 + 2 C_{11}C_{12}+C_{12}^2 + 8C_{13}^2 -2C_{11}C_{33} -2C_{12} C_{33} + C_{33}^2}$.
	
	\item Trigonal $\rightarrow$ Trigonal lattice \newline  
	(6 elastic constants; 4 Kelvin moduli)
\begin{equation}
		\centering
		\left[\bm{\tilde{C}}_{\alpha\beta}\right]\rightarrow\left[
		\begin{array}{cccccc}
			{C_{11}} & {C_{12}} & {C_{13}} & {C_{14}} & {0} & {0} \\
			{C_{12}} & {C_{11}} & {C_{13}} & {-C_{14}} & {0} & {0} \\
			{C_{13}} & {C_{13}} & {C_{33}} & {0} & {0} & {0} \\
			{C_{14}} & {-C_{14}} & {0} & {C_{44}} & {0} & {0} \\
			{0} & {0} & {0} & {0} & {C_{44}} & {\sqrt{2}C_{14}} \\
			{0} & {0} & {0} & {0} & {\sqrt{2}C_{14}} & {C_{11}-C_{12}} \\
		\end{array}
		\right]\,.
		\label{tab:TriCij}
\end{equation}
	Some authors \cite{nye1985physical, Mouhat2014, ZHANG2017403} distinguish a new Trigonal II class with seven, not six, distinct constants.
	However, this is not new class and by a simple transformation can be reduced to the class above \cite{Cowin1995}.
	
	$C_{11} > \left|C_{12}\right|$ \& $C_{44}>0$ \& $2C_{13}^2<C_{33}(C_{11}+C_{12})$ \& $2C_{14}^2<C_{44}(C_{11}-C_{12})$ or\\
	$\lambda_{I, II}=\frac{1}{2}\left(C_{11}-C_{12}+C_{44}-\Diamond\right)>0$ \& 
	$\lambda_{III, IV}=\frac{1}{2}\left(C_{11}-C_{12}+C_{44}+ \Diamond \right)>0$,
	where $\Diamond=\sqrt{C_{11}^2 + 2 C_{11}C_{12}+C_{12}^2 + 8C_{14}^2 -2C_{11}C_{44} +2C_{12} C_{44} + C_{44}^2}$, \&\\ 
	$\lambda_{V}=\frac{1}{2}\left(C_{11}+C_{12}+C_{33}-\bigstar\right)>0$ \& 
	$\lambda_{VI}=\frac{1}{2}\left(C_{11}+C_{12}+C_{33}+\bigstar\right)>0$,	where 
	$\bigstar=\sqrt{C_{11}^2 + 2 C_{11}C_{12}+C_{12}^2 + 8C_{13}^2 -2C_{11}C_{33} -2C_{12} C_{33} + C_{33}^2}$.
	
	\item Tetragonal $\rightarrow$ Tetragonal lattice \newline  
	(6 elastic constants; 5 Kelvin moduli)
\begin{equation}
		\centering
		\left[\bm{\tilde{C}}_{\alpha\beta}\right]\rightarrow\left[
		\begin{array}{cccccc}
			{C_{11}} & {C_{12}} & {C_{13}} & {0} & {0} & {0} \\
			{C_{12}} & {C_{11}} & {C_{13}} & {0} & {0} & {0} \\
			{C_{13}} & {C_{13}} & {C_{33}} & {0} & {0} & {0} \\
			{0} & {0} & {0} & {C_{44}} & {0} & {0} \\
			{0} & {0} & {0} & {0} & {C_{44}} & {0} \\
			{0} & {0} & {0} & {0} & {0} & {C_{66}} \\
		\end{array}
		\right]\,.
		\label{tab:TetrCij}
\end{equation}
	As above some authors \cite{nye1985physical, Mouhat2014, ZHANG2017403} identify also a new Tetragonal II class with seven, not six, distinct constants and again, this is not new class and can be simply transformed to the class above \cite{Cowin1995}.
	
	$C_{11} > \left|C_{12}\right|$ \& $2C_{13}^2<C_{33}(C_{11}+C_{12})$ \& $C_{44}>0$ \& $C_{66}>0$ or\\
	$\lambda_{I, II}=C_{44}>0$ \& $\lambda_{III}=(C_{11}-C_{12})>0$ \& $\lambda_{IV}=C_{66}>0$ \&\\ 
	$\lambda_{V}=\frac{1}{2}\left(C_{11}+C_{22}+C_{33}-\bigstar\right)>0$ \& 
	$\lambda_{VI}=\frac{1}{2}\left(C_{11}+C_{22}+C_{33}+ \bigstar \right)>0$, where $\bigstar=\sqrt{C_{11}^2 + 2 C_{11}C_{12}+C_{12}^2 + 8C_{13}^2 -2C_{11}C_{33} -2C_{12} C_{33} + C_{33}^2}$.
	
	\item Orthotropic $\rightarrow$ Orthorhombic lattice \newline  
	(9 elastic constants; 6 Kelvin moduli)
\begin{equation}
		\centering
		\left[\bm{\tilde{C}}_{\alpha\beta}\right]\rightarrow\left[
		\begin{array}{cccccc}
			{C_{11}} & {C_{12}} & {C_{13}} & {0} & {0} & {0} \\
			{C_{12}} & {C_{22}} & {C_{23}} & {0} & {0} & {0} \\
			{C_{13}} & {C_{23}} & {C_{33}} & {0} & {0} & {0} \\
			{0} & {0} & {0} & {C_{44}} & {0} & {0} \\
			{0} & {0} & {0} & {0} & {C_{55}} & {0} \\
			{0} & {0} & {0} & {0} & {0}      & {C_{66}} \\
		\end{array}
		\right]\,.
		\label{tab:OrtCij}
\end{equation}
	
	$C_{11} >0$ \& $C_{11}C_{22} > C_{12}^2$ \& 
	$C_{11}C_{22}C_{33} + 2C_{12}C_{13}C_{23} - C_{11}C_{23}^2 - C_{22}C_{13}^2 - C_{33}C_{12}^2 > 0$ \& $C_{44} >0$ \& $C_{55} >0$ \& $C_{66} >0$ or $\lambda_{I}=C_{44}>0$ \& $\lambda_{II}=C_{55}>0$ \& $\lambda_{III}=C_{66}>0$ \& $\lambda_{IV}= Root_I|\blacktriangle|>0$ \& $\lambda_{V}=Root_{II}|\blacktriangle|>0$ \& $\lambda_{VI}=Root_{III}|\blacktriangle|>0$, where $\blacktriangle = C_{11}C_{22}C_{33} + 2C_{12}C_{13}C_{23} - C_{11}C_{23}^2 - C_{22}C_{13}^2 - C_{33}C_{12}^2$ (Roots calculated e.g. from the Cardano formula \cite{Itskov07})\\.
	
	\item Monoclinic $\rightarrow$ Monoclinic lattice \newline  
	(13 elastic constants; 6 Kelvin moduli)
\begin{equation}
		\centering
		\left[\bm{\tilde{C}}_{\alpha\beta}\right]\rightarrow\left[
		\begin{array}{cccccc}
			{C_{11}} & {C_{12}} & {C_{13}} & {C_{14}} & {0} & {0} \\
			{C_{12}} & {C_{22}} & {C_{23}} & {C_{24}} & {0} & {0} \\
			{C_{13}} & {C_{23}} & {C_{33}} & {C_{34}} & {0} & {0} \\
			{C_{14}} & {C_{24}} & {C_{34}} & {C_{44}} & {0} & {0} \\
			{0} & {0} & {0} & {0} & {C_{55}} & {C_{56}} \\
			{0} & {0} & {0} & {0} & {C_{56}} & {C_{66}} \\
		\end{array}
		\right]\,.
		\label{tab:MonoCij}
\end{equation}
	
	$\lambda_{i}>0$ (all six eigenvalues of $\bm{\tilde{C}}_{\alpha\beta}$)
	
	\item Triclinic $\rightarrow$ Triclinic lattice \newline  
	(21 elastic constants; 6 Kelvin moduli)
	
\begin{equation}
		\centering
		\left[\bm{\tilde{C}}_{\alpha\beta}\right]\rightarrow\left[
		\begin{array}{cccccc}
			{C_{11}} & {C_{12}} & {C_{13}} & {C_{14}} & {C_{15}} & {C_{16}} \\
			{C_{12}} & {C_{22}} & {C_{23}} & {C_{24}} & {C_{25}} & {C_{26}} \\
			{C_{13}} & {C_{23}} & {C_{33}} & {C_{34}} & {C_{35}} & {C_{36}} \\
			{C_{14}} & {C_{24}} & {C_{34}} & {C_{44}} & {C_{45}} & {C_{46}} \\
			{C_{15}} & {C_{25}} & {C_{35}} & {C_{45}} & {C_{55}} & {C_{56}} \\
			{C_{16}} & {C_{26}} & {C_{36}} & {C_{46}} & {C_{56}} & {C_{66}} \\
		\end{array}
		\right]\,.
		\label{tab:ACij}
\end{equation}

	  $\lambda_{i}>0$ (all six eigenvalues of $\bm{\tilde{C}}_{\alpha\beta}$)
	
\end{enumerate}

In two-dimensional space, there are five different cell lattice types: 
\begin{enumerate}[I.]
	\item Oblique (parallelogram) (a$\neq$b, $\measuredangle\neq$90$^{\circ}$) \label{itm:1}, \label{itm:o}
	\item Rectangular (a$\neq$b, $\measuredangle$=90$^{\circ}$) \label{itm:2},\label{itm:r}
	\item Centered rectangular or diamond (a$\neq$b, $\measuredangle$=90$^{\circ}$) \label{itm:3},\label{itm:cr}
	\item Square (a=b, $\measuredangle$=90$^{\circ}$) \label{itm:4},\label{itm:s}
	\item Rhombic or hexagonal (a=b, $\measuredangle$=120$^{\circ}$) \label{itm:5}.\label{itm:h}
	\label{lt}
\end{enumerate}


Necessary and sufficient elastic stability conditions, also called Born stability
conditions, in various unstressed 3D crystal systems are gathered in \cite{Mouhat2014}, but from my best knowledge, there is no such work for 2D crystal systems. Poza \cite{Mazdziarz2019} "{Comment on 'The Computational 2D Materials Database: high-throughput modeling and discovery of atomically thin crystals'}" tez unstressed
}
\comm{In general, the unstressed crystalline structure is stable with no external loads and in the harmonic approximation, if and only if two independent conditions are fulfilled:
\begin{enumerate}[1.]
	\item All its phonon modes have positive frequencies $\bm{\omega}$ for all wave vectors $\bm{q}$ (dynamical stability):
	\begin{equation}\label{ds}
		\bm{\omega^2(q)}>0,
	\end{equation}
	\item The strain energy density function, given by the quadratic form (\ref{se}), is always
	positive (elastic stability):
	\begin{equation}\label{es}
		U(\bm{\varepsilon})>0, \forall{\bm{\varepsilon}}\neq 0.
	\end{equation}
\end{enumerate}

It is worth pointing out that some authors incorrectly identify elastic stability (\ref{es}) with dynamic stability (\ref{ds}) for the long wave limit (i.e. for vanishing wavevectors $\bm{q}\rightarrow$0) \cite{Grimvall2012, Rehak2012}. In the mathematical elasticity this phonon condition is called strong ellipticity and does not imply positive definiteness of the strain energy density function (\ref{se}), but the opposite implication occurs \cite{hetnarski2010mathematical}. Strong ellipticity admits the existence of a negative bulk modulus $B$ and Poisson's ratio $\nu$ only limited by condition $\notin [\frac{1}{2},1]$, which seems unphysical.

It would be quite difficult to check the positive definiteness of the quadratic form (\ref{es}) directly and it can, therefore, be replaced by one of the equivalent easier conditions \cite{strang2016introduction}:
\begin{enumerate}[1.]
	\item All the leading principal minors of the tensor $\bm{\tilde{C}}$ in (\ref{eqn:Hook3dOrthonormal2}) and (\ref{eqn:HookOrthonormal2}) (determinants of
	its upper-left k by k submatrix) are positive (Sylvester’s criterion),
	
	\item All pivots (the diagonal of the triangle after elimination) of the tensor $\bm{\tilde{C}}$ in \emph{orthonormal} notation (\ref{eqn:Hook3dOrthonormal2}) and (\ref{eqn:HookOrthonormal2}) are positive,
	
	\item All eigenvalues of the tensor $\bm{\tilde{C}}$ in \emph{orthonormal} notation (\ref{eqn:Hook3dOrthonormal2}) and (\ref{eqn:HookOrthonormal2}), called Kelvin moduli, are positive.
\end{enumerate}

}



\comm
{
After this theoretical introduction we can give the form of elastic stiffness tensor $\bm{{C}}$ in the \emph{orthonormal} notation (\ref{eqn:HookOrthonormal2}) and the necessary and sufficient elastic stability conditions (\ref{es}) for all four classes of symmetry for 2D hyperelastic materials.

\begin{enumerate}[1.]
	\item Full symmetry (isotropy) $\rightarrow$ Hexagonal lattice {(\ref{itm:h})} \newline  
	(2 elastic constants; 2 Kelvin moduli)
	\label{itm:iso}
\begin{equation}
		\centering
		\left[\bm{\tilde{C}}_{\alpha\beta}\right]
		\rightarrow\left[
		\begin{array}{ccc}
			{{C}_{11}} & {{C}_{12}} & 0 \\
			{{C}_{12}} & {{C}_{11}} & 0 \\
			0 & 0 & C_{11}-C_{12} \\
		\end{array}
		\right], 
		\label{eqn:Hisotropy}
\end{equation}
	$C_{11}>0$ \& $C_{11} > \left|C_{12}\right|$ or
	$\lambda_I=(C_{11} + C_{12})>0$ \& $\lambda_{II}=(C_{11} - C_{12})>0$.
	\item Symmetry of a square, (tetragonal)$\rightarrow$ Square lattice \newline
	(3 elastic constants; 3 Kelvin moduli)
	\label{itm:sq}
\begin{equation}
		\centering
		\left[\bm{\tilde{C}}_{\alpha\beta}\right]
		\rightarrow\left[
		\begin{array}{ccc}
			{{C}_{11}} & {{C}_{12}} & 0 \\
			{{C}_{12}} & {{C}_{11}} & 0 \\
			0 & 0 & C_{33} \\
		\end{array}
		\right], 
		\label{eqn:Htetragonal}
\end{equation}
	$C_{11}>0$ \& $C_{33}>0$ \& $C_{11} > \left|C_{12}\right|$  or
	$\lambda_I=(C_{11} + C_{12})>0$ \& $\lambda_{II}=(C_{11} - C_{12})>0$ \& $\lambda_{III}=C_{33}>0$.
	\item Symmetry of a rectangle, (orthotropy)$\rightarrow$ Rectangular \& Centered rectangular lattice \newline
	(4 elastic constants; 3 Kelvin moduli)
	\label{itm:rec}
\begin{equation}
		\centering
		\left[\bm{\tilde{C}}_{\alpha\beta}\right]
		\rightarrow\left[
		\begin{array}{ccc}
			{{C}_{11}} & {{C}_{12}} & 0 \\
			{{C}_{12}} & {{C}_{22}} & 0 \\
			0 & 0 & C_{33} \\
		\end{array}
		\right], 
		\label{eqn:Horthotropy}
\end{equation}
	$C_{11}>0$ \& $C_{33}>0$ \& $C_{11}C_{22} > {C^2_{12}} $ or
	$\lambda_I=\frac{1}{2} \left(C_{11} + C_{22}+\sqrt{4C^2_{12}-(C_{11}-C_{22})^2}\right)>0$ \& $\lambda_{II}=\frac{1}{2} \left(C_{11} + C_{22}-\sqrt{4C^2_{12}-(C_{11}-C_{22})^2}\right)>0$ \& $\lambda_{III}=C_{33}>0$. 
	\item No symmetry (anisotropy) $\rightarrow$ Oblique lattice \newline
	(6 elastic constants; 3 Kelvin moduli)
	\label{itm:aniso}
\begin{equation}
		\centering
		\left[\bm{\tilde{C}}_{\alpha\beta}\right]
		\rightarrow\left[
		\begin{array}{ccc}
			{{C}_{11}} & {{C}_{12}} & {{C}_{13}} \\
			{{C}_{12}} & {{C}_{22}} & {{C}_{23}} \\
			{{C}_{13}} & {{C}_{23}} & {{C}_{33}} \\
		\end{array}
		\right], 
		\label{eqn:Hanisotropy}
\end{equation}
	$C_{11}>0$ \& $C_{11}C_{22} > {C^2_{12}}$ \& det($\bm{\tilde{C}}_{\alpha\beta}$)$ > 0 $ or $\lambda_I>0$ \& $\lambda_{II}>0$ \& $\lambda_{III}>0$ (e.g. from the Cardano formula \cite{Itskov07}).
\end{enumerate}

Warunek det(${C}_{\alpha\beta}$) dla zapisu Voigta i Normalnego nie jest rownowazny i może dawać spurious positive definiteness

The problem can arise if we find $C_{13}$ and/or  $C_{23}$ other than zero: it is hard
to say, in this case, if there is no symmetry at all or, possibly, we have chosen a
wrong axis \cite{Blinowski1996}. To avoid this it is recommended to check for all crystals the most general stability condition for anisotropy (\ref{eqn:Hanisotropy}). 

The above considerations are not only of a general nature, selected examples of erroneous stiffness tensors and incorrectly verified elastic stability can be found in the Computational 2D Materials Database (C2DB). \label{sec:C2DB} \newline
	As it was written earlier, crystal symmetry implies symmetries of its physical properties, and hence the symmetries of tensors, e.g. the stiffness tensor. The conditions for elastic stability were given in equations (\ref{eqn:Hisotropy}--\ref{eqn:Hanisotropy}).\newline
	For example, we can find in the C2DB database:
	\begin{itemize}
		\item \textbf{Au$_2$O$_2$}: https://cmrdb.fysik.dtu.dk/c2db/row/Au2O2-GaS-NM\newline
		Space group:P-6m2, $C_{11}$=86.93\,N/m, $C_{22}$=87.90\,N/m and $C_{12}$=103.62\,N/m\newline
		Because it is a Hexagonal lattice {(\ref{itm:h})} the stiffness tensor $\boldsymbol{{c}}$ must be isotropic (\ref{itm:iso}) and  $C_{11}$ must be equal to $C_{22}$. Although all calculated elastic constants are positive, the crystal is not elastically stable because not all required stability conditions (Eq.\ref{eqn:Hisotropy}) are fulfilled. 
		\item \textbf{Ta$_2$Se$_2$}: https://cmrdb.fysik.dtu.dk/c2db/row/Ta2Se2-GaS-FM\newline
		Space group:P-6m2, $C_{11}$=75.15\,N/m, $C_{22}$=75.81\,N/m and $C_{12}$=-45.67\,N/m\newline
		Because it is a Hexagonal lattice {(\ref{itm:h})} the stiffness tensor $\boldsymbol{{c}}$ must be isotropic (\ref{itm:iso}) and  $C_{11}$ must be equal to $C_{22}$. Although calculated elastic constant $C_{12}$ is negative, the crystal is elastically stable because all mandatory stability conditions (Eq.\ref{eqn:Hisotropy}) are satisfied. 
		\item \textbf{Re$_2$O$_2$}: https://cmrdb.fysik.dtu.dk/c2db/row/Re2O2-FeSe-NM\newline
		Space group:P4/nmm, $C_{11}$=17.70\,N/m, $C_{22}$=16.18\,N/m and $C_{12}$=239.42\,N/m\newline
		Because it is a Square lattice {(\ref{itm:s})} the stiffness tensor $\boldsymbol{{c}}$ must have symmetry of a square (\ref{itm:sq}) and  $C_{11}$ must be equal to $C_{22}$ (the difference here is more than 9\%). Although all calculated elastic constants are positive, the crystal is not elastically stable because not all stability requirements (Eq.\ref{eqn:Htetragonal}) are met.  
	\end{itemize} 
}

\section{Internal stability criterion for deformed and stressed lattice}
\label{sec:Iscds}

There are different internal stability criteria encountered in literature based on various tangent or incremental stiffness tensors \cite{Clayton2012}. In the present work, an internal stability criterion with respect to strain increments conjugated to Cauchy stress is applied \cite{Wang1993,Wang1995, Morris2000}. The choice of this criterion is dictated by the fact that it reproduces well the observed physical instabilities and \comm{is the most stringent among the others proposed} is among the two most stringent of the other proposed \cite{Clayton2014}.
Incremental symmetrized tangent modulus $\mathbb{L}$ \cite{Morris2000} is defined by

\begin{equation}\label{bs2}
	L_{ijkl}=C_{ijkl} + \frac{1}{2} \left(\sigma_{ik}\delta_{jl} + \sigma_{il}\delta_{jk} + \sigma_{jk}\delta_{il} + \sigma_{jl}\delta_{ik} - \sigma_{ij}\delta_{kl} - \sigma_{kl}\delta_{ij} \right)\rightarrow \mathbb{L}=\mathbb{C}+\mathbb{H},
\end{equation}
\comm{
\begin{equation}\label{bs}
	L_{ijkl}=D_{ijkl} + H_{ijkl}\rightarrow \mathbb{B}=\mathbb{D}+\mathbb{H},
\end{equation}
\begin{equation}\label{hs}
	H_{ijkl}=\frac{1}{2} \left(\sigma_{ik}\delta_{jl} + \sigma_{il}\delta_{jk} + \sigma_{jk}\delta_{il} + \sigma_{jl}\delta_{ik} - \sigma_{ij}\delta_{kl} - \sigma_{kl}\delta_{ij} \right) ,
\end{equation} }
where 
$\mathbb{C}$ is the {fourth-order} anisotropic stiffness tensor calculated in the current (deformed, stressed) configuration chosen as reference configuration (stiffness tensor $\mathbb{C}$ in Eq.\ref{hl} is calculated in the stress-free configuration), $\bm{\sigma}$ is the second-order Cauchy stress tensor and $\bm{\delta}$ is the Kronecker delta tensor ($i,j,k$=1,2,3 for 3D and $i,j,k$=1,2 for 2D problems). The tensors $\mathbb{C}$, $\mathbb{H}$ and $\mathbb{L}$  possess minor and major symmetry. 

The necessary condition of internal stability requires positive definiteness of the quadratic form
\begin{equation}\label{bse}
	 L_{ijkl}\delta\varepsilon_{ij}\delta\varepsilon_{kl}\geq 0 \rightarrow \bm{\delta\varepsilon}\mathbb{L}\bm{\delta\varepsilon} \geq 0,
\end{equation}
for all incremental strains $\forall{\bm{\delta\varepsilon}}\neq 0$.

Using the \emph{orthonormal} notation described in more detail in Sec.~\ref{sec:srtn}, we can map the fourth-order tensors $\mathbb{C}$, $\mathbb{H}$ and $\mathbb{L}$ in Eq.\ref{bs2} to tensorially equivalent symmetric tensors of order two.

In 3D:

\begin{equation}
	\centering
	\mathbb{L}=\mathbb{C}+\mathbb{H}\rightarrow\left[\bm{\tilde{L}}_{\alpha\beta}\right]=\left[\bm{\tilde{C}}_{\alpha\beta}\right]+\left[\bm{\tilde{H}}_{\alpha\beta}\right], 
	\label{eqn:BCH3d}
\end{equation}

\begin{equation}
	\centering
	\left[\bm{\tilde{C}}_{\alpha\beta}\right]\rightarrow
	\left[
		\begin{array}{cccccc}
			{C_{1111}} & {C_{1122}} & {C_{1133}} & \sqrt{2}{C_{1123}} & \sqrt{2}{C_{1113}} & \sqrt{2}{C_{1112}}\\
			{C_{1122}} & {C_{2222}} & {C_{2233}} & \sqrt{2}{C_{2223}} & \sqrt{2}{C_{2213}} & \sqrt{2}{C_{2212}}\\
			{C_{1133}} & {C_{2233}} & {C_{3333}} & \sqrt{2}{C_{3323}} & \sqrt{2}{C_{3313}} & \sqrt{2}{C_{3312}}\\ 
			\sqrt{2}{C_{1112}} & \sqrt{2}{C_{2212}} & \sqrt{2}{C_{3323}} & 2{C_{2323}} &  2{C_{2313}} &  2{C_{2312}}\\
			\sqrt{2}{C_{1113}} & \sqrt{2}{C_{2213}} & \sqrt{2}{C_{3313}} & 2{C_{2313}} & 2{C_{1313}} & 2{C_{1312}}\\
			\sqrt{2}{C_{1112}} & \sqrt{2}{C_{2212}} & \sqrt{2}{C_{3312}} & 2{C_{2312}} & 2{C_{1312}} & 2{C_{1212}}\\
		\end{array}
	\right]\,,
	\label{tab:ACcijkl}
\end{equation}

and

\begin{equation}\label{h3d}
	\centering
	\left[\bm{\tilde{H}}_{\alpha\beta}\right]
	\rightarrow\left[
	\begin{array}{cccccc}
		\sigma_{11} & \frac{-\sigma_{11}-\sigma_{22}}{2} & \frac{-\sigma_{11}-\sigma_{33}}{2}  & -\frac{\sigma_{23}}{\sqrt{2}} & \frac{\sigma_{13}}{\sqrt{2}} & \frac{\sigma_{12}}{\sqrt{2}} \\
		\frac{-\sigma_{11}-\sigma_{22}}{2} & \sigma_{22} & \frac{-\sigma_{22}-\sigma_{33}}{2}  & \frac{\sigma_{23}}{\sqrt{2}} & -\frac{\sigma_{13}}{\sqrt{2}} & \frac{\sigma_{12}}{\sqrt{2}} \\
		\frac{-\sigma_{11}-\sigma_{33}}{2}  & \frac{-\sigma_{22}-\sigma_{33}}{2}  & \sigma_{33} & \frac{\sigma_{23}}{\sqrt{2}} & \frac{\sigma_{13}}{\sqrt{2}} & -\frac{\sigma_{12}}{\sqrt{2}} \\
		-\frac{\sigma_{23}}{\sqrt{2}} & \frac{\sigma_{23}}{\sqrt{2}} & \frac{\sigma_{23}}{\sqrt{2}} & \sigma_{22}+\sigma_{33} & \sigma_{12} & \sigma_{13} \\
		\frac{\sigma_{13}}{\sqrt{2}} & -\frac{\sigma_{13}}{\sqrt{2}} & \frac{\sigma_{13}}{\sqrt{2}} & \sigma_{12} & \sigma_{11}+\sigma_{33} & \sigma_{23} \\
		\frac{\sigma_{12}}{\sqrt{2}} & \frac{\sigma_{12}}{\sqrt{2}} & -\frac{\sigma_{12}}{\sqrt{2}} & \sigma_{13} & \sigma_{23} & \sigma_{11}+\sigma_{22} \\
	\end{array}
	\right]\,,
\end{equation}

or

\begin{equation}
	\centering
	\left[\bm{\tilde{C}}_{\alpha\beta}\right]\rightarrow
	\left[
	\begin{array}{cccccc}
		{C_{11}} & {C_{12}} & {C_{13}} & {C_{14}} & {C_{15}} & {C_{16}} \\
		{C_{12}} & {C_{22}} & {C_{23}} & {C_{24}} & {C_{25}} & {C_{26}} \\
		{C_{13}} & {C_{23}} & {C_{33}} & {C_{34}} & {C_{35}} & {C_{36}} \\
		{C_{14}} & {C_{24}} & {C_{34}} & {C_{44}} & {C_{45}} & {C_{46}} \\
		{C_{15}} & {C_{25}} & {C_{35}} & {C_{45}} & {C_{55}} & {C_{56}} \\
		{C_{16}} & {C_{26}} & {C_{36}} & {C_{46}} & {C_{56}} & {C_{66}} \\
	\end{array}
	\right]\,,
	\label{tab:ACcij}
\end{equation}

and

\begin{equation}\label{h3d1}
	\centering
	\left[\bm{\tilde{H}}_{\alpha\beta}\right]
	\rightarrow\left[
	\begin{array}{cccccc}
		\sigma_{1} & \frac{-\sigma_{1}-\sigma_{2}}{2} & \frac{-\sigma_{1}-\sigma_{3}}{2}  & -\frac{\sigma_{4}}{{2}} & \frac{\sigma_{5}}{{2}} & \frac{\sigma_{6}}{{2}} \\
		\frac{-\sigma_{1}-\sigma_{2}}{2} & \sigma_{2} & \frac{-\sigma_{2}-\sigma_{3}}{2}  & \frac{\sigma_{4}}{{2}} & -\frac{\sigma_{5}}{{2}} & \frac{\sigma_{6}}{{2}} \\
		\frac{-\sigma_{1}-\sigma_{3}}{2}  & \frac{-\sigma_{2}-\sigma_{3}}{2}  & \sigma_{3} & \frac{\sigma_{4}}{{2}} & \frac{\sigma_{5}}{{2}} & -\frac{\sigma_{6}}{{2}} \\
		-\frac{\sigma_{4}}{{2}} & \frac{\sigma_{4}}{{2}} & \frac{\sigma_{4}}{{2}} & \sigma_{2}+\sigma_{3} & \frac{\sigma_{6}}{\sqrt{2}} & \frac{\sigma_{5}}{\sqrt{2}} \\
		\frac{\sigma_{5}}{{2}} & -\frac{\sigma_{5}}{{2}} & \frac{\sigma_{5}}{{2}} & \frac{\sigma_{6}}{\sqrt{2}} & \sigma_{1}+\sigma_{3} & \frac{\sigma_{4}}{\sqrt{2}} \\
		\frac{\sigma_{6}}{{2}} & \frac{\sigma_{6}}{{2}} & -\frac{\sigma_{6}}{{2}} & \frac{\sigma_{5}}{\sqrt{2}} & \frac{\sigma_{4}}{\sqrt{2}} & \sigma_{1}+\sigma_{2} \\
	\end{array}
	\right]\,.
\end{equation}\\

In 2D:

\begin{equation}
	\centering
	\left[\bm{\tilde{L}}_{\alpha\beta}\right]=\left[\bm{\tilde{C}}_{\alpha\beta}\right]+\left[\bm{\tilde{H}}_{\alpha\beta}\right], 
	\label{eqn:BCH2d}
\end{equation}

\begin{equation}
	\centering
	\left[\bm{\tilde{C}}_{\alpha\beta}\right]
	\rightarrow\left[
	\begin{array}{ccc}
		{C_{1111}} & {C_{1122}} & \sqrt{2}{C_{1112}} \\
		{C_{1122}} & {C_{2222}} & \sqrt{2}{C_{2212}} \\ 
		\sqrt{2}{C_{1112}} & \sqrt{2}{C_{2212}} & 2{C_{1212}} \\
	\end{array}
	\right], 
	\label{eqn:Hcanisotropy1}
\end{equation}

and

\begin{equation}\label{h2d}
	\centering
	\left[\bm{\tilde{H}}_{\alpha\beta}\right]
	\rightarrow\left[
	\begin{array}{ccc}
		\sigma_{11} & \frac{-\sigma_{11}-\sigma_{22}}{2} &  \frac{\sigma_{12}}{\sqrt{2}} \\
		\frac{-\sigma_{11}-\sigma_{22}}{2} & \sigma_{22}  &\frac{\sigma_{12}}{\sqrt{2}} \\
		\frac{\sigma_{12}}{\sqrt{2}} & \frac{\sigma_{12}}{\sqrt{2}} & \sigma_{11}+\sigma_{22} \\
	\end{array}
	\right],
\end{equation}

or

\begin{equation}
	\centering
	\left[\bm{\tilde{C}}_{\alpha\beta}\right]
	\rightarrow\left[
	\begin{array}{ccc}
		{{C}_{11}} & {{C}_{12}} & {{C}_{13}} \\
		{{C}_{12}} & {{C}_{22}} & {{C}_{23}} \\
		{{C}_{13}} & {{C}_{23}} & {{C}_{33}} \\
	\end{array}
	\right], 
	\label{eqn:Hcanisotropy}
\end{equation}

and

\begin{equation}\label{h2d1}
	\centering
	\left[\bm{\tilde{H}}_{\alpha\beta}\right]
	\rightarrow\left[
	\begin{array}{ccc}
		\sigma_{1} & \frac{-\sigma_{1}-\sigma_{2}}{2} &  \frac{\sigma_{3}}{{2}} \\
		\frac{-\sigma_{1}-\sigma_{2}}{2} & \sigma_{2}  &\frac{\sigma_{3}}{{2}} \\
		\frac{\sigma_{3}}{{2}} & \frac{\sigma_{3}}{{2}} & \sigma_{1}+\sigma_{2} \\
	\end{array}
	\right].
\end{equation}

Having already mapped the $\bm{\tilde{L}}_{\alpha\beta}$ tensors, we need to determine the Kelvin moduli, i.e. their eigenvalues, for them. For a 2D problem, Eq.\ref{eqn:BCH2d}, this can be done analytically using the so-called Cardano formulas \cite{Itskov07}. For 3D, except for special cases for the 6$\times$6 tensors, Eq.\ref{eqn:BCH3d}, it is necessary to use numerical procedures using Mathematica \cite{Mathematica}, Python \cite{Python3_2009}, Julia \cite{Julia_2017}, or similar. The positivity of all Kelvin moduli guarantees the positive definiteness of the quadratic form in Eq.\ref{bse} and, consequently, mechanical stability (the tensors $\bm{\tilde{C}}_{\alpha\beta}$, $\bm{\tilde{H}}_{\alpha\beta}$ are symmetric and hence also $\bm{\tilde{L}}_{\alpha\beta}$, so Kelvin moduli are real). 

\section{Conclusions}
\label{sec:con}

As Ludwig Boltzmann is said to have said, "Nothing is more practical than a good theory". The present work is somewhat in this spirit and boils down to a simple recipe on how to algorithmically check the mechanical stability of an arbitrarily loaded material with arbitrary symmetry. {Whether it's a 3D or 2D material, the stiffness tensor in the current configuration can be calculated \emph{ab~initio} or atomistically, and if it's not a stress-free configuration, the Cauchy stress as well.} \comm{Whether it's a 3D or 2D material, {should be calculated} calculate for it \emph{ab~initio} or atomistically the stiffness tensor in the current configuration and, if it's not a stress-free configuration, the Cauchy stress as well.} \comm{Use} {It is important to use} the \emph{orthonormal} notation from Sec.\ref{sec:srtn} to represent the calculated quantities in terms of the corresponding second-order tensors $\bm{\tilde{C}}$, $\bm{\tilde{H}}$ and $\bm{\tilde{L}}$ from Eqs.\ref{eqn:BCH3d}-\ref{h2d1}. {For the tensors of order two represented in this way, their eigenvalues, called Kelvin moduli, can be determined numerically.} The positivity of these moduli indicates the mechanical stability of the material, structure, crystal analyzed.

\section{Supplementary material}
\label{sec:ss}
Mathematica notebook that allows mechanical stability analysis for crystals, stress-free and stressed, of arbitrary symmetry under arbitrary loads is available online at \comm{\href{run:./Mechanical stability.nb}}\href{https://doi.org/10.6084/m9.figshare.29356658}{Supplementary material}\cite{Mazdziarz2025b}.
\section*{ACKNOWLEDGMENTS}
This work was partially supported by the National Science Centre (NCN -- Poland) Research Projects: 2020/37/B/ST8/03907 and 2021/43/B/ST8/03207.

\appendix
\label{sec:App}

\section{NiAl-Stiffness tensors}
\label{sec:AppA}

\comm
{W Appendix daje tensor sztywnosci miedzi za mojej pracy ale dla 2 róznych orientacji, macierz obrotów 6x6 \cite{MEHRABADI1990} str 24 i scicomp.stackexchange.com/questions/35600/4th-order-tensor-rotation-sources-to-refer
i widac ze dla \ref{tab:NiAl2Cij} mamy 6, dla dla \ref{tab:NiAl3Cij} mamy 9 roznych stalych sprezystosci a nie jak w \ref{tab:NiAl1Cij} tylko 3


Stiffness tensor for cooper written in \emph{orthonormal} notation \ref{eqn:Hook3dOrthonormal} for crystal orientation {X=[100] Y=[010] Z=[001]} was taken from the paper \cite{Kowalczyk18}, calculated using molecular statics (MS) \cite{Plimpton1995} and the Embedded Atom Model (EAM) \cite{Mishin2001}. 

\begin{equation}
	\centering
	\left[\bm{\tilde{C}}_{\alpha\beta}\right]\rightarrow\left[
	\begin{array}{cccccc}
		169.909 & 122.632 & 122.632 & {0} & {0} & {0} \\
		122.632 & 169.909 & 122.632 & {0} & {0} & {0} \\
		122.632 & 122.632 & 169.909 & {0} & {0} & {0} \\
		{0} & {0} & {0} & 152.381 & {0} & {0} \\
		{0} & {0} & {0} & {0} & 152.381 & {0} \\
		{0} & {0} & {0} & {0} & {0} & 152.381 \\
	\end{array}
	\right],
	\label{tab:Cu1Cij}
\end{equation}

the same stiffness tensor for cooper but determined for crystal orientation \mbox{X=[110]} \mbox{Y=[1-10]} \mbox{Z=[001]} (the orthogonal rotation tensor in 6-dimensional space can be found in \cite{MEHRABADI1990})

\begin{equation}
	\centering
	\left[\bm{\tilde{C}}_{\alpha\beta}\right]\rightarrow\left[
	\begin{array}{cccccc}
		222.461 & 70.0801 & 122.632 & {0} & {0} & {0} \\
		70.0801 & 222.461 & 122.632 & {0} & {0} & {0} \\
		122.632 & 122.632 & 169.909 & {0} & {0} & {0} \\
		{0} & {0} & {0} & 152.381 & {0} & {0} \\
		{0} & {0} & {0} & {0} & 152.381 & {0} \\
		{0} & {0} & {0} & {0} & {0} & 47.2772 \\
	\end{array}
	\right],
	\label{tab:Cu2Cij}
\end{equation}

and the same stiffness tensor for cooper but determined for crystal orientation {X=[111] Y=[1-10] Z=[11-2]}

\begin{equation}
	\centering
	\left[\bm{\tilde{C}}_{\alpha\beta}\right]\rightarrow\left[
	\begin{array}{cccccc}
		234.139 & 81.7583 & 99.2755 & -8.25771 & -8.25771 & 16.5154 \\
		81.7583 & 234.139 & 99.2755 & -8.25771 & -8.25771 & 16.5154 \\
		99.2755 & 99.2755 & 216.622 & 16.5154 & 16.5154 & -33.0308 \\
		-8.25771 & -8.25771 & 16.5154 & 105.668 & -46.7127 & -11.6782 \\
		-8.25771 & -8.25771 & 16.5154 & -46.7127 & 105.668 & -11.6782 \\
		16.5154 & 16.5154 & -33.0308 & -11.6782 & -11.6782 & 70.6336 \\
	\end{array}
	\right]\,.
	\label{tab:Cu3Cij}
\end{equation}
}

Stiffness tensor for stress-free B2 NiAl written in \emph{orthonormal} notation \ref{eqn:Hook3dOrthonormal} for crystal orientation {X=[100] Y=[010] Z=[001]} was taken from the paper \cite{MAZDZIARZ2024109953}, calculated using  molecular statics (MS) approach in LAMMPS \cite{Plimpton1995} and the Embedded Atom Model (EAM) \cite{Purja2009}. 

\begin{equation}
	\centering
	\left[\bm{\tilde{C}}_{\alpha\beta}\right]\rightarrow\left[
	\begin{array}{cccccc}
		190.868 & 142.908 & 142.908 & 0. & 0. & 0. \\
		142.908 & 190.868 & 142.908 & 0. & 0. & 0. \\
		142.908 & 142.908 & 190.868 & 0. & 0. & 0. \\
		0. & 0. & 0. & 242.971 & 0. & 0. \\
		0. & 0. & 0. & 0. & 242.971 & 0. \\
		0. & 0. & 0. & 0. & 0. & 242.971 \\
	\end{array}
	\right],
	\label{tab:NiAl1Cij}
\end{equation}

the same stiffness tensor for B2 NiAl but determined for crystal orientation \mbox{X=[110]} \mbox{Y=[1-10]} \mbox{Z=[001]} (the \emph{orthogonal} rotation tensor in 6-dimensional space can be found in \cite{MEHRABADI1990})

\begin{equation}
	\centering
	\left[\bm{\tilde{C}}_{\alpha\beta}\right]\rightarrow\left[
	\begin{array}{cccccc}
		288.374 & 45.402 & 142.908 & 0. & 0. & 0. \\
		45.402 & 288.374 & 142.908 & 0. & 0. & 0. \\
		142.908 & 142.908 & 190.868 & 0. & 0. & 0. \\
		0. & 0. & 0. & 242.971 & 0. & 0. \\
		0. & 0. & 0. & 0. & 242.971 & 0. \\
		0. & 0. & 0. & 0. & 0. & 47.960 \\
	\end{array}
	\right],
	\label{tab:NiAl2Cij}
\end{equation}

and the same stiffness tensor for B2 NiAl but determined for crystal orientation {X=[111] Y=[1-10] Z=[11-2]}

\begin{equation}
	\centering
	\left[\bm{\tilde{C}}_{\alpha\beta}\right]\rightarrow\left[
	\begin{array}{cccccc}
		320.876 & 77.904 & 77.904 & 0. & 0. & 0. \\
		77.904 & 288.374 & 110.406 & 0. & 0. & -65.004 \\
		77.904 & 110.406 & 288.374 & 0. & 0. & 65.004 \\
		0. & 0. & 0. & 177.968 & 91.929 & 0. \\
		0. & 0. & 0. & 91.929 & 112.964 & 0. \\
		0. & -65.004 & 65.004 & 0. & 0. & 112.964 \\
	\end{array}
	\right]\,.
	\label{tab:NiAl3Cij}
\end{equation}

Depending on the orientation of the crystal, we have here 3 \ref{tab:NiAl1Cij}, 6 \ref{tab:NiAl2Cij} and 9 \ref{tab:NiAl3Cij} distinct elastic constants, respectively. As can be seen, the pattern of the stiffness tensor also changes, but the number of Kelvin moduli remains constant. For all 3 orientations, the Kelvin moduli are identical and are as follows $\lambda_i$=(476.684, 242.971, 242.971, 242.971, 47.96, 47.96), $i$=1,\dots,6.

\section{NiAl-Stability analysis}
\label{sec:AppB}


Let us consider the application of a homogeneous deformation to a monocrystal, in accordance with the Cauchy-Born rule (hypothesis) \cite{Clayton2011}.

The ordering of indexes in the symmetric strain and stress tensor is here as follows : 1$\rightarrow$11, 2$\rightarrow$22, 3$\rightarrow$33, 4$\rightarrow$23, 5$\rightarrow$13 and 6$\rightarrow$12, whereas in LAMMPS \cite{Plimpton1995} is assumed 1$\rightarrow$11, 2$\rightarrow$22, 3$\rightarrow$33, 4$\rightarrow$12, 5$\rightarrow$13, 6$\rightarrow$23, and in the Cartesian coordinate system, 1 is X, 2 is Y, and 3 is Z.


Consider a biaxial strain state, that is, one in which the deformation gradient \textbf{F} can be written in the following form

\begin{equation}
	\centering
	\left[\mathbf{{F}}_{ij}\right]
	\rightarrow\left[
	\begin{array}{ccc}
		{\alpha} & 0 & 0 \\
		0 & {\alpha} & 0 \\ 
		0 & 0 & 1 \\
	\end{array}
	\right], 
	\label{eqn:Fbs}
\end{equation}

where $\alpha$ is stretch ratio, {for $\alpha$>1 we have tension and for 0<$\alpha$<1 compression}. The Lagrangian finite strain tensor, also called the Green strain tensor is defined as $\mathbf{E=\frac{1}{2}(F^{T}F-I)}$. Linearization of the Green strain tensor yields the infinitesimal strain tensor $\mathbf{\varepsilon=\frac{1}{2}(F^{T}+F-2I)}$, also called linear strain tensor, or small strain tensor \cite{Hackett2018}.

For the B2 NiAl crystal oriented {in the Cartesian coordinate system} such that {X=[100] Y=[010] Z=[001]} using molecular statics calculations we are looking for such $\alpha$ that in the current, deformed configuration $\bm{\tilde{C}}_{\alpha\beta}$ in Eq.\ref{tab:ACcijkl} becomes singular.


\begin{equation}
	\centering
	\left[\bm{\tilde{C}}_{\alpha\beta}\right]\rightarrow\left[
	\begin{array}{cccccc}
		12.351 & 12.375 & 5.225 & 0. & 0. & 0. \\
		12.375 & 12.351 & 5.225 & 0. & 0. & 0. \\
		5.225 & 5.225 & 54.415 & 0. & 0. & 0. \\
		0. & 0. & 0. & 61.726 & 0. & 0. \\
		0. & 0. & 0. & 0. & 61.726 & 0. \\
		0. & 0. & 0. & 0. & 0. & 84.722 \\
	\end{array}
	\right]\,.
	\label{tab:NiAl4Cij}
\end{equation}

For {biaxial tension} and $\alpha$ equal to 1.15365, the Kelvin moduli are as follows $\lambda^{\bm{\tilde{C}}}_i$=(84.722, 61.726, 61.726, 56.1524, 22.989, -0.024), $i$=1,\dots,6 and $\bm{\tilde{C}}_{\alpha\beta}$ becomes singular.

We perform analogous calculations, but now require that $\bm{\tilde{L}}_{\alpha\beta}$ in Eq.\ref{eqn:BCH3d} become singular.  


\begin{equation}
	\centering
	\left[\bm{\tilde{C}}_{\alpha\beta}\right]\rightarrow\left[
	\begin{array}{cccccc}
		1.179 & 9.579 & 3.781 & 0. & 0. & 0. \\
		9.579 & 1.179 & 3.781 & 0. & 0. & 0. \\
		3.781 & 3.781 & 53.815 & 0. & 0. & 0. \\
		0. & 0. & 0. & 59.011 & 0. & 0. \\
		0. & 0. & 0. & 0. & 59.011 & 0. \\
		0. & 0. & 0. & 0. & 0. & 79.308 \\
	\end{array}
	\right]\,.
	\label{tab:NiAl5Cij}
\end{equation}

For $\alpha$ equal to 1.15454, the Kelvin moduli are as follows $\lambda^{\bm{\tilde{C}}}_i$=(79.308, 59.011, 59.011, 54.469, 10.105, -8.400), $i$=1,\dots,6. As can be seen, $\bm{\tilde{C}}_{\alpha\beta}$ alone is also singular here. 


Since this is a deformed crystal, the stresses are not zero and are $\sigma_{11}$ = 27.060, $\sigma_{22}$ = 27.060, and $\sigma_{33}$ = 20.585\,GPa, respectively. For this stress state, the $\bm{\tilde{H}}_{\alpha\beta}$ tensor in Eq.\ref{h3d1} is

\begin{equation}
	\centering
	\left[\bm{\tilde{H}}_{\alpha\beta}\right]\rightarrow\left[
	\begin{array}{cccccc}
		27.060 & -27.060 & -23.823 & 0. & 0. & 0. \\
		-27.060 & 27.060 & -23.823 & 0. & 0. & 0. \\
		-23.823 & -23.823 & 20.585 & 0. & 0. & 0. \\
		0. & 0. & 0. & 47.645 & 0. & 0. \\
		0. & 0. & 0. & 0. & 47.645 & 0. \\
		0. & 0. & 0. & 0. & 0. & 54.120 \\
	\end{array}
	\right]\,,
	\label{tab:NiAlHij}
\end{equation}

and the resulting $\bm{\tilde{L}}_{\alpha\beta}$ becomes

\begin{equation}
	\centering
	\left[\bm{\tilde{L}}_{\alpha\beta}\right]\rightarrow\left[
	\begin{array}{cccccc}
		28.239 & -17.481 & -20.042 & 0. & 0. & 0. \\
		-17.481 & 28.239 & -20.042 & 0. & 0. & 0. \\
		-20.042 & -20.042 & 74.400 & 0. & 0. & 0. \\
		0. & 0. & 0. & 106.656 & 0. & 0. \\
		0. & 0. & 0. & 0. & 106.656 & 0. \\
		0. & 0. & 0. & 0. & 0. & 133.427 \\
	\end{array}
	\right]\,,
	\label{tab:NiAlBij}
\end{equation}

and the Kelvin moduli are as follows $\lambda^{\bm{\tilde{L}}}_i$=(133.427, 106.656, 106.656, 85.193, 45.720, -0.034), $i$=1,\dots,6 and $\bm{\tilde{L}}_{\alpha\beta}$ becomes singular.
\comm{The calculations confirmed earlier observations that the singularity condition for $\bm{\tilde{L}}_{\alpha\beta}$ is more stringent than that for $\bm{\tilde{C}}_{\alpha\beta}$ and reduces the allowable deformation for biaxial strain from $\alpha$=1.15365 to $\alpha$=1.15454 while the crystal still remains mechanically stable.} 
{We see that for biaxial tension the singularity condition for $\bm{\tilde{C}}_{\alpha\beta}$ ($\alpha$=1.15365) is somewhat more stringent than that for $\bm{\tilde{L}}_{\alpha\beta}$ ($\alpha$=1.15454), while the crystal still remains mechanically stable.}

{Similar to the case of biaxial tension, biaxial compression will now be analyzed. Again, we are looking for such $\alpha$ that in the current, deformed configuration $\bm{\tilde{C}}_{\alpha\beta}$ in Eq.\ref{tab:ACcijkl} becomes singular.

\begin{equation}
	\centering
	\left[\bm{\tilde{C}}_{\alpha\beta}\right]\rightarrow\left[
	\begin{array}{cccccc}
		1137.230 & 813.821 & 908.399 & 0. & 0. & 0. \\
		813.821 & 1137.230 & 908.399 & 0. & 0. & 0. \\
		908.399 & 908.399 & 845.906 & 0. & 0. & 0. \\
		0. & 0. & 0. & 399.541 & 0. & 0. \\
		0. & 0. & 0. & 0. & 399.541 & 0. \\
		0. & 0. & 0. & 0. & 0. & 1.875 \\
	\end{array}
	\right]\,.
	\label{tab:NiAl4cCij}
\end{equation}

For {biaxial compression} and $\alpha$ equal to 0.7985, the Kelvin moduli are as follows $\lambda^{\bm{\tilde{C}}}_i$=(2796.950, 399.541, 399.541, 323.411, 1.875, 0.011), $i$=1,\dots,6 and $\bm{\tilde{C}}_{\alpha\beta}$ becomes singular.

We perform analogous calculations, but now require that $\bm{\tilde{L}}_{\alpha\beta}$ in Eq.\ref{eqn:BCH3d} become singular.  


\begin{equation}
	\centering
	\left[\bm{\tilde{C}}_{\alpha\beta}\right]\rightarrow\left[
	\begin{array}{cccccc}
		621.924 & 280.913 & 336.130 & 0. & 0. & 0. \\
		280.913 & 621.924 & 336.130 & 0. & 0. & 0. \\
		336.130 & 336.130 & 372.138 & 0. & 0. & 0. \\
		0. & 0. & 0. & 431.175 & 0. & 0. \\
		0. & 0. & 0. & 0. & 431.175 & 0. \\
		0. & 0. & 0. & 0. & 0. & 253.649 \\
	\end{array}
	\right]\,.
	\label{tab:NiAl5cCij}
\end{equation}

For $\alpha$ equal to 0.9087, the Kelvin moduli are as follows $\lambda^{\bm{\tilde{C}}}_i$=(1181.890, 431.175, 431.175, 341.011, 253.649, 93.083), $i$=1,\dots,6. As can be seen, $\bm{\tilde{C}}_{\alpha\beta}$ alone is not singular here. 


Since this is a deformed crystal, the stresses are not zero and are $\sigma_{11}$ = -59.567, $\sigma_{22}$ = -59.567, and $\sigma_{33}$ = -41.191\,GPa, respectively. For this stress state, the $\bm{\tilde{H}}_{\alpha\beta}$ tensor in Eq.\ref{h3d1} is

\begin{equation}
	\centering
	\left[\bm{\tilde{H}}_{\alpha\beta}\right]\rightarrow\left[
	\begin{array}{cccccc}
		-59.567 & 59.567 & 50.379 & 0. & 0. & 0. \\
		59.567 & -59.567 & 50.379 & 0. & 0. & 0. \\
		50.379 & 50.379 & -41.191 & 0. & 0. & 0. \\
		0. & 0. & 0. & -100.758 & 0. & 0. \\
		0. & 0. & 0. & 0. & -100.758 & 0. \\
		0. & 0. & 0. & 0. & 0. & -119.134 \\
	\end{array}
	\right]\,,
	\label{tab:NiAlcHij}
\end{equation}

and the resulting $\bm{\tilde{L}}_{\alpha\beta}$ becomes

\begin{equation}
	\centering
	\left[\bm{\tilde{L}}_{\alpha\beta}\right]\rightarrow\left[
	\begin{array}{cccccc}
		562.357 & 340.480 & 386.508 & 0. & 0. & 0. \\
		340.480 & 562.357 & 386.508 & 0. & 0. & 0. \\
		386.508 & 386.508 & 330.947 & 0. & 0. & 0. \\
		0. & 0. & 0. & 330.417 & 0. & 0. \\
		0. & 0. & 0. & 0. & 330.417 & 0. \\
		0. & 0. & 0. & 0. & 0. & 134.515 \\
	\end{array}
	\right]\,,
	\label{tab:NiAlcBij}
\end{equation}

and the Kelvin moduli are as follows $\lambda^{\bm{\tilde{L}}}_i$=(1233.770, 330.417, 330.417, 221.876, 134.515, 0.011), $i$=1,\dots,6 and $\bm{\tilde{L}}_{\alpha\beta}$ becomes singular.
We see that for biaxial compression the singularity condition for $\bm{\tilde{L}}_{\alpha\beta}$ ($\alpha$=0.9087) is much more stringent than that for $\bm{\tilde{C}}_{\alpha\beta}$ ($\alpha$=0.7985), while the crystal still remains mechanically stable. Thus, \comm{one should check} it is recommended to check the condition on $\bm{\tilde{C}}_{\alpha\beta}$ as well as on $\bm{\tilde{L}}_{\alpha\beta}$. Similar observations were made in \cite{Clayton2014}, where the behavior of a certain isotropic material subjected to different loads was analyzed.}

The Mathematica notebook \cite{Mathematica} that allows to analyze the mechanical stability of crystals, stress-free and stressed, of arbitrary symmetry subjected to arbitrary loads is available in the Supplementary material \ref{sec:ss} \comm{\href{run:./Mechanical stability.nb}{Supplementary Material}}.

\section{3D-Explicitly written out mechanical stability conditions}
\label{sec:App3D}

The following representations of stiffness tensor are given with respect to the symmetry axes, in a canonical base, with standard lattice vectors \cite{ZHANG2017403}. \comm{In molecular or \emph{ab~initio} calculations, we don't always use conventional computational cells. Sometimes it is more convenient to convert non-orthogonal cells, such as hexagonal or trigonal cells, into orthogonal cells \cite{MAZDZIARZ2024109953}.
}

\begin{enumerate}[1.]
		
		\item Cubic \& Isotropy $\rightarrow$ Cubic lattice {(\ref{tab:3DS})} \newline  
		(3 \& 2  elastic constants; 3 \&2 Kelvin moduli)
		\begin{equation}
			\centering
			\left[\bm{\tilde{C}}_{\alpha\beta}\right]\rightarrow\left[
			\begin{array}{cccccc}
				{C_{11}}  & {C_{12}} & {C_{12}} & {0}      & {0}      & {0} \\
				{C_{12}}  & {C_{11}} & {C_{12}} & {0}      & {0}      & {0} \\
				{C_{12}}  & {C_{12}} & {C_{11}} & {0}      & {0}      & {0} \\
				{0}       & {0}      & {0}      & {C_{44}} & {0}      & {0}    \\
				{0}       & {0}      & {0}      & {0}      & {C_{44}} & {0}    \\
				{0}       & {0}      & {0}      & {0}      & {0}      & {C_{44}}\\
			\end{array}
			\right]\,,
			\label{tab:CubicCij}
		\end{equation}
		
		(For Isotropy $C_{44}=C_{11}-C_{12}$)
		
		$C_{11}-C_{12}>0$ \& $C_{11}+2C_{12}>0$ \& $C_{44}>0$ or\\
		$\lambda_{I, II, III}=C_{44}>0$ \& $\lambda_{IV, V}=(C_{11}-C_{12})>0$ \& $\lambda_{VI}=(C_{11}+2C_{12})>0$.

		\item Transverse Isotropy $\rightarrow$ Hexagonal lattice {(\ref{tab:3DS})} \newline  
		(5 elastic constants; 4 Kelvin moduli)
		\begin{equation}
			\centering
			\left[\bm{\tilde{C}}_{\alpha\beta}\right]\rightarrow\left[
			\begin{array}{cccccc}
				{C_{11}} & {C_{12}} & {C_{13}} & {0} & {0} & {0} \\
				{C_{12}} & {C_{11}} & {C_{13}} & {0} & {0} & {0} \\
				{C_{13}} & {C_{13}} & {C_{33}} & {0} & {0} & {0} \\
				{0} & {0} & {0} & {C_{44}} & {0} & {0} \\
				{0} & {0} & {0} & {0} & {C_{44}} & {0} \\
				{0} & {0} & {0} & {0} & {0} & {C_{11}-C_{12}} \\
			\end{array}
			\right]\,,
			\label{tab:TICij}
		\end{equation}
		
		$C_{11} > \left|C_{12}\right|$ \& $2C_{13}^2>C_{33}(C_{11}+C_{12})$ \& $C_{44}>0$ \& $(C_{11}-C_{12})>0$ or\\
		$\lambda_{I, II}=C_{44}>0$ \& $\lambda_{III, IV}=(C_{11}-C_{12})>0$ \& \\ 
		$\lambda_{V}=\frac{1}{2}\left(C_{11}+C_{22}+C_{33}-\bigstar\right)>0$ \& 
		$\lambda_{VI}=\frac{1}{2}\left(C_{11}+C_{22}+C_{33}+\bigstar\right)>0$, where $\bigstar=\sqrt{C_{11}^2 + 2 C_{11}C_{12}+C_{12}^2 + 8C_{13}^2 -2C_{11}C_{33} -2C_{12} C_{33} + C_{33}^2}$.
		
		\item Trigonal $\rightarrow$ Trigonal lattice {(\ref{tab:3DS})} \newline  
		(6 elastic constants; 4 Kelvin moduli)
		\begin{equation}
			\centering
			\left[\bm{\tilde{C}}_{\alpha\beta}\right]\rightarrow\left[
			\begin{array}{cccccc}
				{C_{11}} & {C_{12}} & {C_{13}} & {C_{14}} & {0} & {0} \\
				{C_{12}} & {C_{11}} & {C_{13}} & {-C_{14}} & {0} & {0} \\
				{C_{13}} & {C_{13}} & {C_{33}} & {0} & {0} & {0} \\
				{C_{14}} & {-C_{14}} & {0} & {C_{44}} & {0} & {0} \\
				{0} & {0} & {0} & {0} & {C_{44}} & {\sqrt{2}C_{14}} \\
				{0} & {0} & {0} & {0} & {\sqrt{2}C_{14}} & {C_{11}-C_{12}} \\
			\end{array}
			\right]\,.
			\label{tab:TriCij}
		\end{equation}
		Some authors \cite{nye1985physical, Mouhat2014, ZHANG2017403} distinguish a new Trigonal II class with seven, not six, distinct constants.
		However, this is not new class and by a simple transformation can be reduced to the class above \cite{Cowin1995}.
		
		$C_{11} > \left|C_{12}\right|$ \& $C_{44}>0$ \& $2C_{13}^2<C_{33}(C_{11}+C_{12})$ \& $2C_{14}^2<C_{44}(C_{11}-C_{12})$ or\\
		$\lambda_{I, II}=\frac{1}{2}\left(C_{11}-C_{12}+C_{44}-\Diamond\right)>0$ \& 
		$\lambda_{III, IV}=\frac{1}{2}\left(C_{11}-C_{12}+C_{44}+ \Diamond \right)>0$,
		where $\Diamond=\sqrt{C_{11}^2 + 2 C_{11}C_{12}+C_{12}^2 + 8C_{14}^2 -2C_{11}C_{44} +2C_{12} C_{44} + C_{44}^2}$, \&\\ 
		$\lambda_{V}=\frac{1}{2}\left(C_{11}+C_{12}+C_{33}-\bigstar\right)>0$ \& 
		$\lambda_{VI}=\frac{1}{2}\left(C_{11}+C_{12}+C_{33}+\bigstar\right)>0$,	where 
		$\bigstar=\sqrt{C_{11}^2 + 2 C_{11}C_{12}+C_{12}^2 + 8C_{13}^2 -2C_{11}C_{33} -2C_{12} C_{33} + C_{33}^2}$.
		
		\item Tetragonal $\rightarrow$ Tetragonal lattice {(\ref{tab:3DS})} \newline  
		(6 elastic constants; 5 Kelvin moduli)
		\begin{equation}
			\centering
			\left[\bm{\tilde{C}}_{\alpha\beta}\right]\rightarrow\left[
			\begin{array}{cccccc}
				{C_{11}} & {C_{12}} & {C_{13}} & {0} & {0} & {0} \\
				{C_{12}} & {C_{11}} & {C_{13}} & {0} & {0} & {0} \\
				{C_{13}} & {C_{13}} & {C_{33}} & {0} & {0} & {0} \\
				{0} & {0} & {0} & {C_{44}} & {0} & {0} \\
				{0} & {0} & {0} & {0} & {C_{44}} & {0} \\
				{0} & {0} & {0} & {0} & {0} & {C_{66}} \\
			\end{array}
			\right]\,.
			\label{tab:TetrCij}
		\end{equation}
		As above some authors \cite{nye1985physical, Mouhat2014, ZHANG2017403} identify also a new Tetragonal II class with seven, not six, distinct constants and again, this is not new class and can be simply transformed to the class above \cite{Cowin1995}.
		
		$C_{11} > \left|C_{12}\right|$ \& $2C_{13}^2<C_{33}(C_{11}+C_{12})$ \& $C_{44}>0$ \& $C_{66}>0$ or\\
		$\lambda_{I, II}=C_{44}>0$ \& $\lambda_{III}=(C_{11}-C_{12})>0$ \& $\lambda_{IV}=C_{66}>0$ \&\\ 
		$\lambda_{V}=\frac{1}{2}\left(C_{11}+C_{22}+C_{33}-\bigstar\right)>0$ \& 
		$\lambda_{VI}=\frac{1}{2}\left(C_{11}+C_{22}+C_{33}+ \bigstar \right)>0$, where $\bigstar=\sqrt{C_{11}^2 + 2 C_{11}C_{12}+C_{12}^2 + 8C_{13}^2 -2C_{11}C_{33} -2C_{12} C_{33} + C_{33}^2}$.
		
		\item Orthotropic $\rightarrow$ Orthorhombic lattice {(\ref{tab:3DS})} \newline  
		(9 elastic constants; 6 Kelvin moduli)
		\begin{equation}
			\centering
			\left[\bm{\tilde{C}}_{\alpha\beta}\right]\rightarrow\left[
			\begin{array}{cccccc}
				{C_{11}} & {C_{12}} & {C_{13}} & {0} & {0} & {0} \\
				{C_{12}} & {C_{22}} & {C_{23}} & {0} & {0} & {0} \\
				{C_{13}} & {C_{23}} & {C_{33}} & {0} & {0} & {0} \\
				{0} & {0} & {0} & {C_{44}} & {0} & {0} \\
				{0} & {0} & {0} & {0} & {C_{55}} & {0} \\
				{0} & {0} & {0} & {0} & {0}      & {C_{66}} \\
			\end{array}
			\right]\,,
			\label{tab:OrtCij}
		\end{equation}
		
		$C_{11} >0$ \& $C_{11}C_{22} > C_{12}^2$ \& 
		$C_{11}C_{22}C_{33} + 2C_{12}C_{13}C_{23} - C_{11}C_{23}^2 - C_{22}C_{13}^2 - C_{33}C_{12}^2 > 0$ \& $C_{44} >0$ \& $C_{55} >0$ \& $C_{66} >0$ or $\lambda_{I}=C_{44}>0$ \& $\lambda_{II}=C_{55}>0$ \& $\lambda_{III}=C_{66}>0$ \& $\lambda_{IV}= Root_I|\blacktriangle|>0$ \& $\lambda_{V}=Root_{II}|\blacktriangle|>0$ \& $\lambda_{VI}=Root_{III}|\blacktriangle|>0$, where $\blacktriangle = C_{11}C_{22}C_{33} + 2C_{12}C_{13}C_{23} - C_{11}C_{23}^2 - C_{22}C_{13}^2 - C_{33}C_{12}^2$ (Roots calculated e.g. from the Cardano formula \cite{Itskov07}).
		
		\item Monoclinic $\rightarrow$ Monoclinic lattice {(\ref{tab:3DS})} \newline  
		(13 elastic constants; 6 Kelvin moduli)
		\begin{equation}
			\centering
			\left[\bm{\tilde{C}}_{\alpha\beta}\right]\rightarrow\left[
			\begin{array}{cccccc}
				{C_{11}} & {C_{12}} & {C_{13}} & {C_{14}} & {0} & {0} \\
				{C_{12}} & {C_{22}} & {C_{23}} & {C_{24}} & {0} & {0} \\
				{C_{13}} & {C_{23}} & {C_{33}} & {C_{34}} & {0} & {0} \\
				{C_{14}} & {C_{24}} & {C_{34}} & {C_{44}} & {0} & {0} \\
				{0} & {0} & {0} & {0} & {C_{55}} & {C_{56}} \\
				{0} & {0} & {0} & {0} & {C_{56}} & {C_{66}} \\
			\end{array}
			\right]\,,
			\label{tab:MonoCij}
		\end{equation}
		
		$\lambda_{i}>0$ (all six eigenvalues of $\bm{\tilde{C}}_{\alpha\beta}$)
		
		\item Triclinic $\rightarrow$ Triclinic lattice {(\ref{tab:3DS})} \newline  
		(21 elastic constants; 6 Kelvin moduli)
		
		\begin{equation}
			\centering
			\left[\bm{\tilde{C}}_{\alpha\beta}\right]\rightarrow\left[
			\begin{array}{cccccc}
				{C_{11}} & {C_{12}} & {C_{13}} & {C_{14}} & {C_{15}} & {C_{16}} \\
				{C_{12}} & {C_{22}} & {C_{23}} & {C_{24}} & {C_{25}} & {C_{26}} \\
				{C_{13}} & {C_{23}} & {C_{33}} & {C_{34}} & {C_{35}} & {C_{36}} \\
				{C_{14}} & {C_{24}} & {C_{34}} & {C_{44}} & {C_{45}} & {C_{46}} \\
				{C_{15}} & {C_{25}} & {C_{35}} & {C_{45}} & {C_{55}} & {C_{56}} \\
				{C_{16}} & {C_{26}} & {C_{36}} & {C_{46}} & {C_{56}} & {C_{66}} \\
			\end{array}
			\right]\,,
			\label{tab:ACij}
		\end{equation}

		$\lambda_{i}>0$ (all six eigenvalues of $\bm{\tilde{C}}_{\alpha\beta}$)
		
	\end{enumerate}

\section{2D-Explicitly written out mechanical stability conditions} 
\label{sec:App2D}

\comm{After this theoretical introduction we can give the form of stiffness tensor $\bm{{C}}$ in the \emph{orthonormal} notation (\ref{eqn:HookOrthonormal2}) and the necessary and sufficient elastic stability conditions (\ref{es2}) for all four classes of symmetry for 2D hyperelastic materials.}

\begin{enumerate}[1.]
		\item Full symmetry (isotropy) $\rightarrow$ Hexagonal lattice {(\ref{tab:2DS})} \newline 
		(2 elastic constants; 2 Kelvin moduli)
		\label{itm:iso}
		\begin{equation}
			\centering
			\left[\bm{\tilde{C}}_{\alpha\beta}\right]
			\rightarrow\left[
			\begin{array}{ccc}
				{{C}_{11}} & {{C}_{12}} & 0 \\
				{{C}_{12}} & {{C}_{11}} & 0 \\
				0 & 0 & C_{11}-C_{12} \\
			\end{array}
			\right], 
			\label{eqn:Hisotropy}
		\end{equation}
		$C_{11}>0$ \& $C_{11} > \left|C_{12}\right|$ or
		$\lambda_I=(C_{11} + C_{12})>0$ \& $\lambda_{II}=(C_{11} - C_{12})>0$.
		\item Symmetry of a square, (tetragonal)$\rightarrow$ Square lattice {(\ref{tab:2DS})} \newline
		(3 elastic constants; 3 Kelvin moduli)
		\label{itm:sq}
		\begin{equation}
			\centering
			\left[\bm{\tilde{C}}_{\alpha\beta}\right]
			\rightarrow\left[
			\begin{array}{ccc}
				{{C}_{11}} & {{C}_{12}} & 0 \\
				{{C}_{12}} & {{C}_{11}} & 0 \\
				0 & 0 & C_{33} \\
			\end{array}
			\right], 
			\label{eqn:Htetragonal}
		\end{equation}
		$C_{11}>0$ \& $C_{33}>0$ \& $C_{11} > \left|C_{12}\right|$  or
		$\lambda_I=(C_{11} + C_{12})>0$ \& $\lambda_{II}=(C_{11} - C_{12})>0$ \& $\lambda_{III}=C_{33}>0$.
		\item Symmetry of a rectangle, (orthotropy)$\rightarrow$ Rectangular \& Centered rectangular lattice {(\ref{tab:2DS})} \newline
		(4 elastic constants; 3 Kelvin moduli)
		\label{itm:rec}
		\begin{equation}
			\centering
			\left[\bm{\tilde{C}}_{\alpha\beta}\right]
			\rightarrow\left[
			\begin{array}{ccc}
				{{C}_{11}} & {{C}_{12}} & 0 \\
				{{C}_{12}} & {{C}_{22}} & 0 \\
				0 & 0 & C_{33} \\
			\end{array}
			\right], 
			\label{eqn:Horthotropy}
		\end{equation}
		$C_{11}>0$ \& $C_{33}>0$ \& $C_{11}C_{22} > {C^2_{12}} $ or
		$\lambda_I=\frac{1}{2} \left(C_{11} + C_{22}+\sqrt{4C^2_{12}-(C_{11}-C_{22})^2}\right)>0$ \& $\lambda_{II}=\frac{1}{2} \left(C_{11} + C_{22}-\sqrt{4C^2_{12}-(C_{11}-C_{22})^2}\right)>0$ \& $\lambda_{III}=C_{33}>0$. 
		\item No symmetry (anisotropy) $\rightarrow$ Oblique lattice {(\ref{tab:2DS})} \newline
		(6 elastic constants; 3 Kelvin moduli)
		\label{itm:aniso}
		\begin{equation}
			\centering
			\left[\bm{\tilde{C}}_{\alpha\beta}\right]
			\rightarrow\left[
			\begin{array}{ccc}
				{{C}_{11}} & {{C}_{12}} & {{C}_{13}} \\
				{{C}_{12}} & {{C}_{22}} & {{C}_{23}} \\
				{{C}_{13}} & {{C}_{23}} & {{C}_{33}} \\
			\end{array}
			\right], 
			\label{eqn:Hanisotropy}
		\end{equation}
		$C_{11}>0$ \& $C_{11}C_{22} > {C^2_{12}}$ \& det($\bm{\tilde{C}}_{\alpha\beta}$)$ > 0 $ or $\lambda_I>0$ \& $\lambda_{II}>0$ \& $\lambda_{III}>0$ (e.g. from the Cardano formula \cite{Itskov07}).
	\end{enumerate}
	
	\comm{Warunek det($\bm{\tilde{C}}_{\alpha\beta}$) dla zapisu Voigta i Normalnego nie jest rownowazny i może dawać spurious positive definiteness
		The problem can arise if we find $C_{13}$ and/or  $C_{23}$ other than zero: it is hard
	to say, in this case, if there is no symmetry at all or, possibly, we have chosen a
	wrong axis \cite{Blinowski1996}. To avoid this it is recommended to check for all crystals the most general stability condition for anisotropy (\ref{eqn:Hanisotropy}). 
	}
	
	If $C_{13}$ and/or $C_{23}$ are non-zero, it is difficult to determine if there is no symmetry or if the axes are incorrect \cite{Blinowski1996}. To avoid this, the most general condition for anisotropy (\ref{eqn:Hanisotropy}) should be checked.

\section{Homogenized isotropic bulk and shear modulus}
\label{sec:AppE}
In the \emph{orthonormal} notation regardless of the choice of axes orientation we get:
\begin{description}
	\item[Voigt averaging]
	\begin{equation}\label{bv}
		\textrm{Bulk modulus,}\quad{B_V}= \frac{1}{9}\left[\left(C_{11}+C_{22}+C_{33}\right)+2\left(C_{12}+C_{13}+C_{23}\right)  \right],
    \end{equation}	
	\begin{equation}\label{gv}	
		\textrm{Shear modulus,}\quad{G_V}= \frac{1}{15}\left[\left(C_{11}+C_{22}+C_{33}\right)-\left(C_{12}+C_{13}+C_{23}\right)+\frac{3}{2}\left(C_{44}+C_{55}+C_{66}\right)  \right].
	\end{equation}
\item[Reuss averaging]
	\begin{equation}\label{br}
		\textrm{Bulk modulus,}\quad{B_R}= \frac{1}{\left[\left(S_{11}+S_{22}+S_{33}\right)+2\left(S_{12}+S_{13}+S_{23}\right)\right]},
	\end{equation}	
	\begin{equation}\label{gr}	
		\textrm{Shear modulus,}\quad{G_R}= \frac{15}{\left[4\left(S_{11}+S_{22}+S_{33}\right)-4\left(S_{12}+S_{13}+S_{23}\right)+6\left(S_{44}+S_{55}+S_{66}\right)\right]},
	\end{equation}
where $S_{ij}$ are the elements of the compliance tensor $\bm{\tilde{S}}=\bm{\tilde{C}}^{-1}$.
\end{description}

\bibliography{References}

\end{document}